\documentclass{emulateapj}
\usepackage{amsmath}
\usepackage{amssymb}
\usepackage{graphicx}
\usepackage{appendix}

\usepackage{morefloats}

\newcommand{\lSect}[1]{{\label{sec:#1}}}
\newcommand{\lFig}[1]{{\label{fig:#1}}}
\newcommand{\lEq}[1]{{\label{eq:#1}}}
\newcommand{\lTab}[1]{{\label{tab:#1}}}
\newcommand{\Msun}{{\ensuremath{M_{\odot}}}}

\newcommand{\Zsun}{\ensuremath{\mathrm{Z}_{\odot}}}

\newcommand{\FIGFF}[2]{{\ref{fig:#2}{#1}}}

\newcommand{\FIG}[2]{{Fig.~\FIGFF{#1}{#2}}}
\newcommand{\Fig}[1]{{\FIG{}{#1}}}

\newcommand{\Sectff}[1]{{\ref{sec:#1}}}
\newcommand{\Sect}[1]{{\S~\Sectff{#1}}}

\newcommand{\Eqref}[1]{{\ref{eq:#1}}}

\newcommand{\eqff}[1]{{\Eqref{#1}}}

\newcommand{\eq}[1]{{equation~\eqff{#1}}}

\newcommand{\Tab}[1]{{Table~\ref{tab:#1}}}

\newcommand{\cp}{\ensuremath{\mathrm{\xi}_{2.5}}}

\begin{document}

\submitted{Submitted to The Astrophysical Journal on 10 October, 2013.}

\title{The Compactness of Presupernova Stellar Cores}

\author{Tuguldur Sukhbold\altaffilmark{1} and
  S. E. Woosley\altaffilmark{1}}

\altaffiltext{1}{Department of Astronomy and Astrophysics, University
  of California, Santa Cruz, CA 95064; sukhbold@ucolick.org}

\begin{abstract}
The success or failure of the neutrino-transport mechanism for
producing a supernova in an evolved massive star is known to be
sensitive not only to the mass of the iron core that collapses, but also to
the density gradient in the silicon and oxygen shells surrounding that
core.  Here we study the systematics of a presupernova core's
``compactness'' \citep{Oco11} as a function of the mass of the star
and the physics used in its calculation. Fine-meshed surveys of
presupernova evolution are calculated for stars from 15 to 65
\Msun. The metallicity and the efficiency of semiconvection and
overshoot mixing are both varied and bare carbon-oxygen cores are
explored as well as full hydrogenic stars. Two different codes, KEPLER
and MESA, are used for the study. A complex interplay of carbon and
oxygen burning, especially in shells, can cause rapid variations in the
compactness for stars of very nearly the same mass.  On larger scales,
the distribution of compactness with main sequence mass is found to be
robustly non-monotonic, implying islands of ``explodability'',
particularly around 8 to 20 \Msun \ and 25 to 30 \Msun. The
carbon-oxygen (CO) core mass of a presupernova star is a better,
though still ambiguous discriminant of its core structure than the
main sequence mass.
\end{abstract}


\section{Introduction}
\lSect{s1}

The compactness of its core is an important structural characteristic
of a presupernova star that affects whether it will explode as a
supernova. A shallow density gradient around the iron core, as
typically exists in more massive stars, implies a higher accretion
rate and ''ram pressure'' \citep{Coo84} surrounding the iron core
during its collapse that must be overcome by neutrino energy
deposition or other energy deposition to turn what is initially an
implosion into an explosion.  The connection between this structure
and the likelihood of explosion has been noted many times
\citep{Bur87,Fry99,Oco11,Ugl12}, but little attention has been given
to explaining just why the compactness has the values that it does, or
why different groups obtain different core structures for
models with similar main sequence mass.

Recently, \citet{Oco11} quantified this compactness in terms of a parameter,
\begin{equation}
\xi_{M}=\frac{M/\Msun}{R(M_{bary}=M)/1000\, {\rm km}}\Big|_{t_{bounce}},
\lEq{compact}
\end{equation}

where M=2.5 \Msun \ was chosen as the relevant mass for quantifying
the density gradient outside of the iron core, which itself typically
has a mass in the range 1.4 to 2.0 \Msun.  In O'Connor et al's
definition, the time, t$_{\rm bounce}$, chosen for evaluating
$\xi_{M}$ is when the core has collapsed to its maximum
(super-nuclear) density. As we shall show (\Sect{s2.3}), for M = 2.5
\Msun, no substantial accuracy is lost if this parameter is evaluated
for the ``presupernova model'' defined by when the collapse speed
first reaches 1000 km s$^{-1}$.  The choice of M = 2.5 \Msun\ is
justified as being larger than the iron core itself, yet deep enough
in the star to sample matter that might accrete, especially in a
failed explosion. 
Since \cp\ is inversely proportional to the 
radius of a fiducial mass outside the iron core, it is small when the density 
around the iron core falls off rapidly with radius and greater when the 
density gradient is shallow. That is, when \cp\ is large, one need go only a 
shorter distance to include more mass.
The systematics of neutrino-powered supernova explosions were explored as
a function of \cp \ by \citet{Oco11} and \citet{Ugl12}, both of whom
found that a small value of \cp\ favored explosion. \citet{Oco11}
plotted \cp\ as a function of main sequence mass for a variety of
surveys due to \citet{Woo02}, \citet{Woo07} and \citet{LC06}, and a
subset of that data is shown in \Fig{f1}. Note especially the
distinctly non-monotonic behavior in the vicinity of 25 \Msun \ and 40
\Msun.

For a polytrope of constant index and composition supported by a
constant fraction ($\beta$) of ideal gas pressure, the compactness
defined by \eq{compact} will actually {\sl decrease} monotonically
with increasing mass when evaluated at a constant central temperature
and small radius.  This is because, for polytropes, the central
temperature and density obey a relation \citep[e.g.][]{Woo02}

\begin{equation}
\frac{T_c^3}{\rho_c} \propto M^2
\lEq{poly}
\end{equation}

For a given $T_c$ the central density thus declines as $1/M^2$ and the
radius required to enclose a given mass increases with M. As will be
discussed in \Sect{s3.1}, \eq{poly} works reasonably well
for massive stars that are burning hydrogen and helium in their
centers. The inner regions of such stars are non-degenerate and can be
well represented by polytropes of constant index.

As massive stars evolve, however, they develop nested
cores with different compositions and entropies. Lighter
stars develop more degenerate cores in their late stages, especially
during oxygen burning. The large gravitational potential at the edge
of these degenerate cores implies a small pressure scale height that
results in a steep decline in the local density
\citep{Fau01,Fau05}. As a result, in heavier stars, the higher entropy
actually leads to more extended configurations and larger values of
\cp. One does not have to to go out as far in the ``envelope'' of the
compact core to encompass 2.5 \Msun, while in lighter stars that mass
is reached farther out. This accounts for a general tendency of the
compactness parameter for presupernova stars to increase with mass,
especially for low metallicity stars where mass loss is less
important.

\begin{figure}
\includegraphics[width=0.35\textwidth,angle=90]{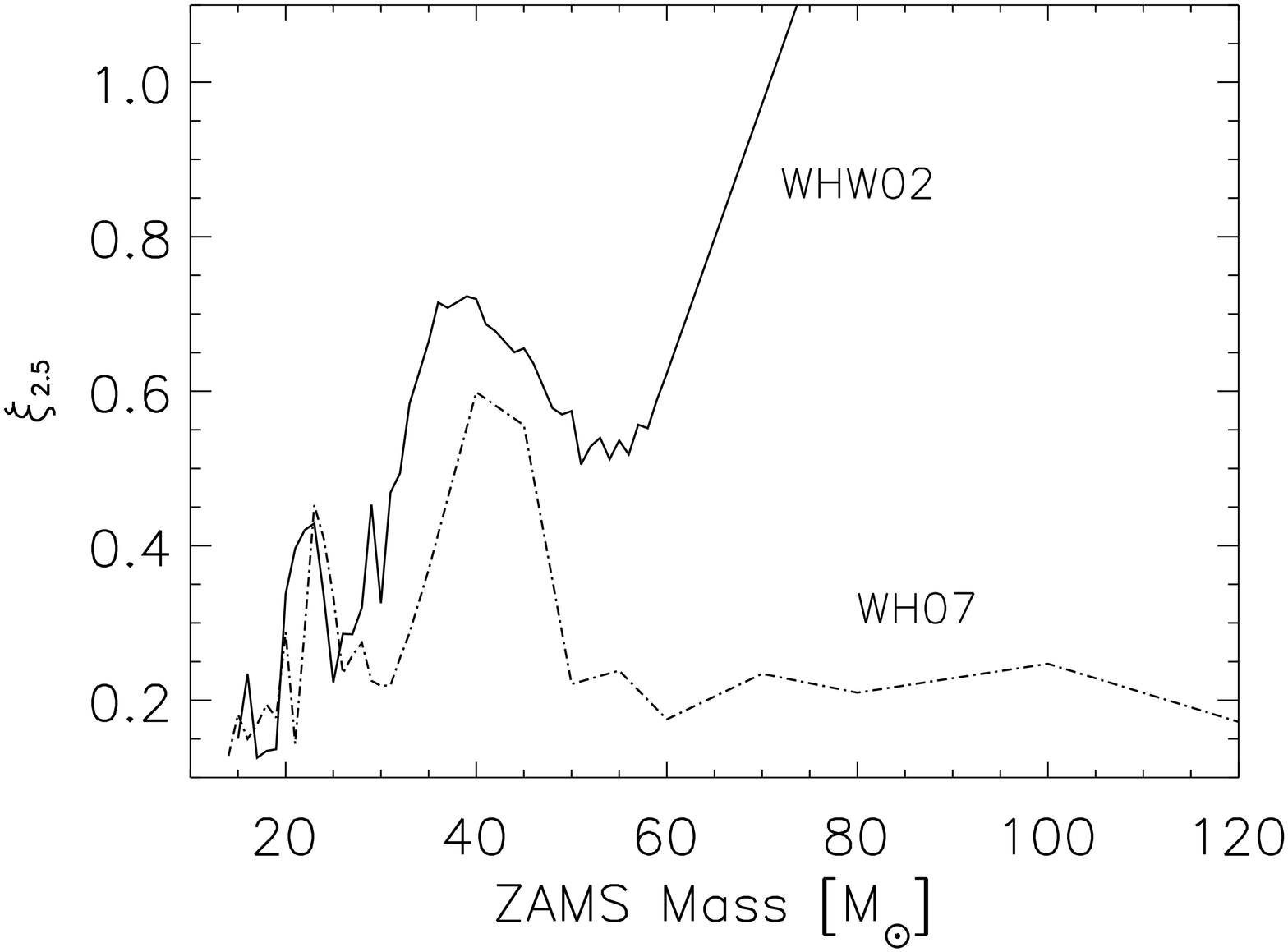}
\caption{The compactness parameter as a function of mass for the very
  low metallicity (10$^{-4}$ \Zsun) models from \citet{Woo02} (thick)
  and for the solar metallicity models from \citet{Woo07}
  (dot-dashed). See also \Fig{f9} of \citet{Oco11}. Note the
  non-monotonic behavior in both sets of models around 25 \Msun and 40
  \Msun. Since the mass loss was ignored in the low metallicity models,
  the mass of the CO core and the compactness continues to increase
  for those models at very high masses. \lFig{f1}}
\end{figure}

\Fig{f1} also shows that the core compactness has significant
non-monotonic behavior above about 20 \Msun, and possibly some fine
structure below 20 \Msun. If this behavior is real and
robust, it would have interesting implications for the explosion
mechanism(s) for massive stars, the masses of their compact remnants,
and stellar nucleosynthesis. To address the ``real and robust'' issue,
it is necessary to understand why this non-monotonic structure exists
and its sensitivity to uncertainties and variations in the stellar
models.

The present study has several parts, some dealing with the systematics
of \cp \ found in new surveys of massive stars, others with the
uncertainties one should assign to those results. In \Sect{s2}, three
new surveys of massive stellar evolution are presented for stars of
solar metallicity, very low metallicity (10$^{-4}$ solar), and solar
metallicity with suppressed mass loss. These new surveys, though using
the same physics as in previous works \citep{Woo07}, are needed to
provide a finer mass grid for examining rapid fluctuations in \cp\ and
to give additional data not archived in previous work. In \Sect{s3},
we address the heart of the matter: why is the compactness a
non-monotonic function and, in some places, almost chaotic function of
main sequence mass. The timing and location of several carbon and
oxygen convective shells are found to play a major role. Because there
are multiple episodes of shell burning, the final compactness can be
quite complex. In \Sect{s4} the sensitivity of these results to
uncertain assumptions in the physics used in two stellar evolution
codes, KEPLER and MESA, is explored. These uncertainties, especially
the treatment of semiconvection and convective overshoot mixing, lead
to large variations in the final CO core mass that emerges for a given
main sequence mass and account for much of the diversity of published
results for presupernova evolution. Since these uncertainties cloud
the interpretation of the compactness plot, a further study is carried
out in \Sect{s5} for bare CO stars using both codes. Provided the
carbon mass fractions at carbon ignition are the same as
in the full star models, the same trends seen previously for the
compactness parameter are also found in these simpler cores, albeit
with an offset due to surface boundary pressure in the full star. Thus
the non-monotonicity of \cp\ can be considered robust and the CO core
mass is a somewhat better indicator of pre-supernova
structure (and to some extent explosion dynamics) than the main
sequence mass. Finally, \Sect{s6} offers some conclusions.

\section{New Surveys of Presupernova Evolution Using the KEPLER CODE}
\lSect{s2}

\subsection{Solar Metallicity Stars}

The models published in \citep{Woo07} were aimed at sampling supernova
nucleosynthesis and light curves for a broad range of non-rotating
stars of solar metallicity. The resolution in mass necessary for that
purpose was not particularly fine, so that study has been repeated
using the same code and physics, but with a finer mass grid. The
prescriptions for nuclear reaction rates, mass loss, opacity,
convective overshoot mixing, and semiconvection are the same as in the
previous survey. For the assumed mass loss rates
  \citep{NdJ90}, the entire hydrogen envelope is lost for stars
with initial masses above 33 \Msun. Heavier stars thus die as
Wolf-Rayet (WR) stars with smaller helium cores than would have
existed had mass loss been neglected. For the mass loss
  rate adopted during the WR-phase \citep{Wel99}, the core shrinks to
the extent that the final compactness becomes small again, thus the
compactness of very heavy massive stars when they die depends upon an
uncertain prescription for mass loss.

\begin{figure}[htb]
\centering
\includegraphics[width=0.35\textwidth,angle=90]{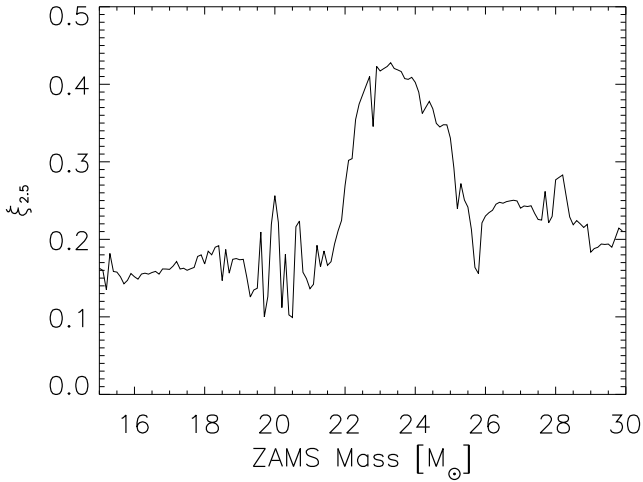}
\caption{The core compactness determined for 151 KEPLER
  presupernova models derived from solar metallicity stars of the
  given masses (the ``S series''). \cp\ is evaluated when the collapse
  speed anywhere in the iron core first reaches 1000 km s$^{-1}$
  (see \Fig{f7} for other choices). Notice especially the low
  values of \cp\ for stars lighter than 22 \Msun, the irregular
  variations from 18 to 22 \Msun, and the subsequent rise to a maximum
  at 23 - 24 \Msun \ followed by a decline. For heavier stars
  additional structure, including a second rise, is found at higher
  masses (\Fig{f3}).\lFig{f2}}
\end{figure}

The new survey (\Fig{f2}) includes non-rotating stars with solar
metallicity and masses in the range 15 to 30 \Msun \ calculated in
increments of 0.1 \Msun using the 1D implicit hydrodynamic code KEPLER
(Table 1). Altogether, 151 models were simulated using the standard
physics (for variations see \Sect{s4}). Nuclear energy
generation was calculated, up to oxygen depletion (central oxygen mass
fraction less than 0.04), using a 19 isotope network \citep{Wea78}. A
quasi-statistical equilibrium network and a nuclear statistical
equilibrium network were used thereafter.  However, a much larger
``adaptive'' network of up to 1500 isotopes was carried in parallel
with the main structure calculation \citep{Rau02} and was used to
calculate changes in the electron mole number, $Y_e$. This was
particularly important during oxygen burning when the
quasi-equilibrium network could not be employed, but substantial weak
interactions were already changing the central structure. This is the
same approach that was used by \citep{Woo07}, but a substantial
improvement over what was done in \citet{Woo02}.  These models are
collectively referred to as the ``S series'' stars - for ``solar''.

In addition to confirming the distinct ''bump'' previously seen around
23 \Msun \ the new study (\Fig{f2}) shows other interesting
features. Fine scale variation of the compactness parameter persists
throughout most of the mass range, but is particularly apparent
between 19 and 21 \Msun. These features are further explored in
\Sect{s3}.

For this series of models, \cp \ is small for main sequence stars
lighter than 22 \Msun. It then rises rapidly to a peak, but declines
again to lower values for main sequence masses between 26 \Msun \ and
30 \Msun. Above about 30 \Msun, results for these solar metallicity
stars are clouded by the effects of mass loss reducing the helium core
mass, but if mass loss is suppressed, the compactness stays large
above 40 \Msun \ with a slight dip around 50 \Msun \ (\Fig{f1} and $\S
2.2$).  These features pose a challenge to the conventional notion
that the difficulty of exploding a star is a monotonically increasing
function of its initial mass and instead imply that models below 22
\Msun \ and between 26 \Msun \ and 30 \Msun \ may be easier to blow-up
than other masses. This could alter how we think about the galactic
chemical evolution \citep{Bro13}.

In order to facilitate the comparison with low metallicity stars where
the mass loss rate may be negligible, another series of solar
metallicity models was calculated for stars above 30 \Msun \ in which
mass loss was artificially suppressed. These are the ''SH models''
covering the mass range of 30 - 60 \Msun \ with varying increments -
31,32,33,35,40,..,60 \Msun (Table 1). This survey was truncated at 60
\Msun \ because the effect of the pulsational pair instability became
noticeable around 70 - 80 \Msun.  The 60 \Msun \ model of the
SH-series has a helium core mass of 27.5 \Msun\ (\Fig{f4}), far
greater than that of any of the solar metallicity stars calculated
with mass loss.

For all runs, the convective time history was recorded and checkpoints
registered at representative points along the evolution including: 1)
helium depletion - when helium reached 1\% by mass in the stellar
center; 2) carbon ignition - when the central temperature was $5
\times 10^8$ K; 3) carbon depletion - when the central carbon mass
fraction fell below 1\%; 4) oxygen ignition - when the central
temperature first reached $1.6 \times 10^9$ K; 5) oxygen depletion -
when the central oxygen abundance declined to 4\%; 6) silicon
depletion - when silicon mass fraction reached 1\% in the center; and
7) presupernova - when any point in the core collapsed faster than 1000 km
s$^{-1}$. Subsequent discussion will refer to these points as
representative of the stellar models at these different times as
He.dep., O.ign., preSN., etc.

\subsection{Low Metallicity Stars}
\lSect{s2.2}

Stars with equal masses on the main sequence, but differing initial
metallicity can have different presupernova structure for a variety of
reasons. Most importantly, metallicity affects the mass loss. If the
amount of mass lost is very low or zero, the presupernova star,
including its helium core, is larger, and that has a dramatic effect
on the compactness (\Fig{f1}). There are also other less dramatic,
but important effects. Low metallicity implies a smaller initial
helium mass fraction (and more hydrogen). The final helium core mass
is sensitive to this and is reduced. Nitrogen has an abundance just
prior to helium ignition that is directly proportional to the initial
metallicity and ``nitrogen burning'' by
$^{14}$N($\alpha,\gamma)^{18}$F($e^+\nu)^{18}$O is an important
exoergic, convective phase in the star's life that precedes helium
burning. Low metallicity affects the energy generation during hydrogen
shell burning by the CNO cycle, and this affects the boundary conditions for
the helium core. Low metallicity also affects the opacity, and the
combined effects of opacity, energy generation, and mass loss
determine if the star is a red supergiant or a blue one when it
dies. This is especially true for the non-rotating stars studied
here. The more compact radiative structure of a blue supergiant
envelope places greater surface boundary pressure on the helium core
therein and can affect its evolution.

To illustrate the systematics of compactness in stars of low
metallicity, we include here a previously unpublished set of
non-rotating presupernova stars by Heger and Woosley with metallicity of
10$^{-4}$ \Zsun. These used the same physics as the S series, but had
an initial composition of 76\% hydrogen and 24\% helium with only a
trace, 10$^{-4}$ solar, of heavier elements. The masses of these stars
were 10 to 95 \Msun, though only a subset, 15 - 65 \Msun, is
considered here. This restricted set, called the U series, contains 86
models with varying mass increments (0.2\Msun between 15 - 25\Msun,
0.5\Msun between 25-35 \Msun, 1\Msun between 35 - 45\Msun \ and 5\Msun
between 45 - 65 \Msun; (Table 1). Other details of these models not
related to their compactness will be published elsewhere.

\begin{figure}[htb]
\centering
\includegraphics[width=0.48\textwidth]{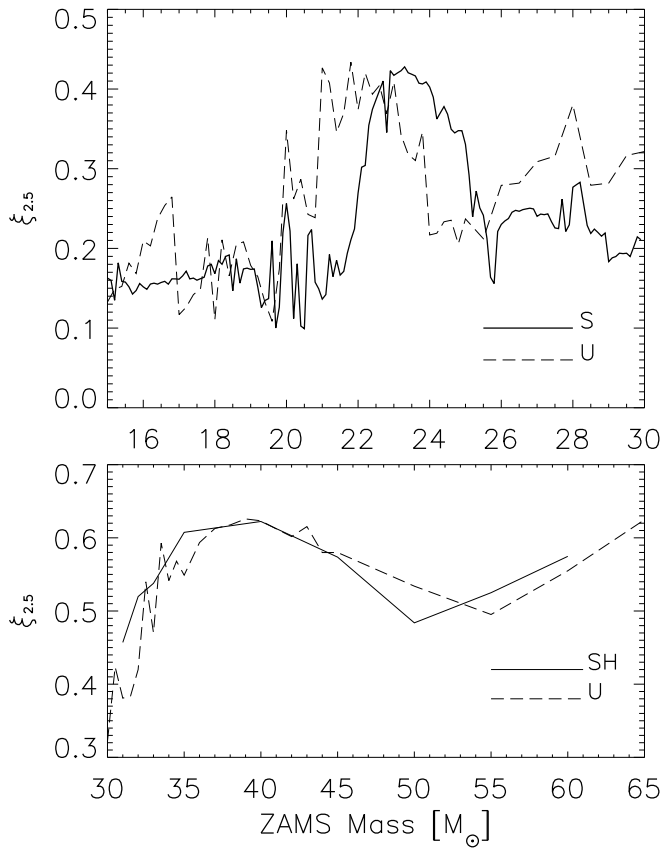}
\caption{Top: the final core compactness parameters evaluated at
  2.5\Msun\ are shown for the S-series (solar metallicity; thick line)
  and U-series (10$^{-4}$ solar metallicity; dashed) stars of mass
  less than 30 \Msun. Notice a slight shift of \cp \ near the first
  "bump".  Bottom: same plot for the U-series (dashed) and SH-series
  (continuous) models for masses over 30 \Msun. \lFig{f3}}
\end{figure}

\begin{figure}[htb]
\centering
\includegraphics[width=0.48\textwidth]{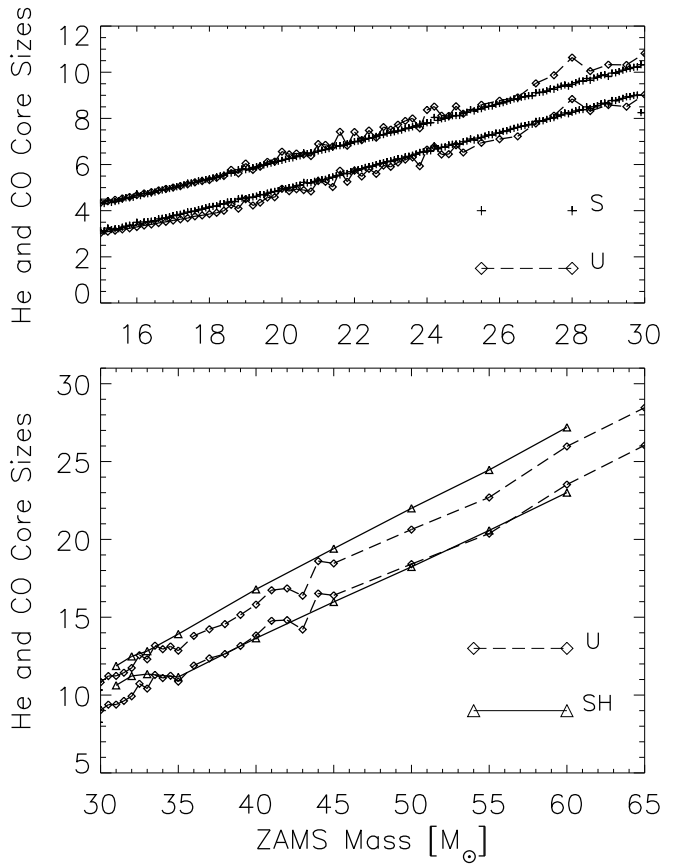}
\caption{Core masses are shown as a function of initial mass for the
  new surveys. The top panel shows the helium (upper curve) and CO
  (lower curve) core masses for the S and U series; the bottom panel
  shows the core masses for the SH and U series above 30 \Msun.  Below
  35 \Msun, the core masses are not very sensitive to metallicity.
  Above 35 \Msun, the helium core is somewhat larger for the higher
  metallicity stars reflecting a more active hydrogen-burning shell
  during helium burning. \lFig{f4}}
\end{figure}

The compactness of these low metallicity models is shown, along with
those from S and SH series, in \Fig{f3}. Without mass loss, a much
larger range of helium and CO-core masses is accessible (Figure
4). One sees a continuation of the overall increase in compactness
parameter with increasing mass all the way up to 65 \Msun, but with
new features. The compactness rises rapidly above 30 \Msun\ to a broad
peak around 40 \Msun \ and then, following a dip at 50 \Msun, resumes
its rise. All of the stars above 30 \Msun, with no mass loss, will
probably be very difficult to explode by any solely neutrino-powered
mechanism.

\begin{figure}[htb]
\centering
\includegraphics[width=0.35\textwidth,angle=90]{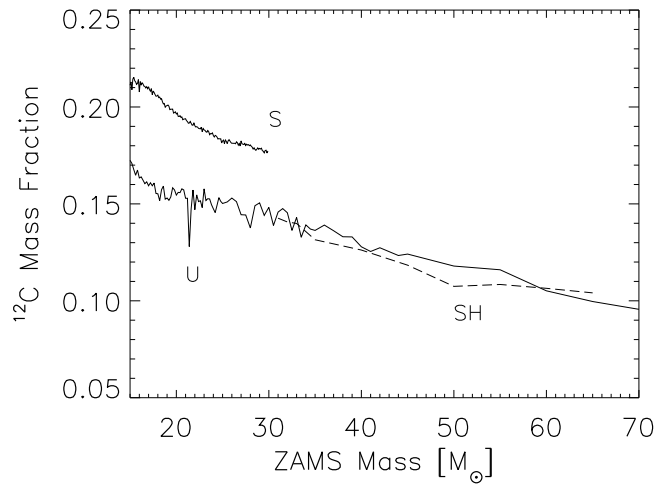}
\caption{The $^{12}C$ mass fraction at the time of core carbon
  ignition is shown for S (thick upper curve), SH (dashed) and U
  (thinner lower curve) stars. Note that the carbon mass fraction is
  smaller in stars that ignore mass loss and that the SH series
  tracks the U series very well. \lFig{f5}}
\end{figure}

Below 30 \Msun, the compactness curve for the U series shows
non-monotonic structure with peaks qualitatively similar to those
observed for the S series in Figure 2, but with an offset of about a
solar mass.  The offset is most pronounced in the range 20 - 25 \Msun
\ where the compactness starts to rise in the U series for a lower
value of main sequence mass.  The main cause of this shift is the lower
mass fraction of carbon produced by helium burning in the U stars
(\Fig{f5}). That value, in turn, affects the critical mass where
carbon ceases to burn convectively in the star's center (\Sect{s3.1}),
shifting it to lower values, which causes the compactness curve to
rise earlier (\Sect{s3}).

\begin{figure}[htb]
\centering
\includegraphics[width=0.35\textwidth,angle=90]{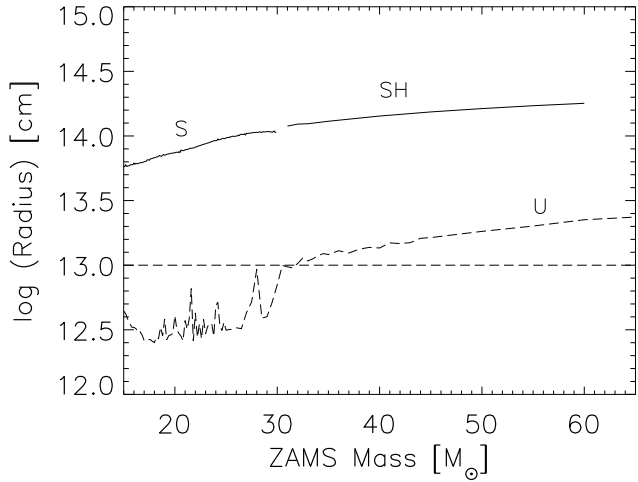}
\caption{The final radii of the presupernova stars are shown from
  different surveys - S (thick), SH (thin) and U (dashed). The dotted
  horizontal line marks the radius of a red supergiant;
  U-stars below about 30 \Msun \ thus end their lives as blue
  supergiants. \lFig{f6} }
\end{figure}

This variable nucleosynthesis of carbon for the S and U stars results
from differing amounts of helium being mixed into the convective
helium core as it grows during the final stages of helium burning. In
fact, the carbon mass fractions for S and U stars in this mass range
are very similar half way through helium burning, but as the helium
mass fraction declines below 10\%, the U stars are more effective at
bringing additional helium in from their outer regions. Each helium
convected down turns a carbon into an oxygen. The U stars have
somewhat higher helium core luminosities at this point and the
structure at the outer edge of the helium convection is also
influenced by surface boundary pressure from the hydrogen envelope
which is different for red and blue supergiants. Below 30 \Msun, the
U-series stars are blue supergiants while the S-series stars are red
ones (\Fig{f6}). The pressure at the edge of the helium core does not
decline as steeply in a blue supergiant and the entropy barriers
inhibiting convection in the outer helium core are slightly
reduced. This makes the growth of the helium convective core
easier. Convective dredge up in a red supergiant could also possibly
reduce the helium core mass, though this effect seems to be small in
the present models.  With rotation, many of these lighter U-stars
would also be red supergiants in the end because of primary nitrogen
production at the hydrogen-helium interface, and the effect would be
diminished. Above 30 \Msun \ the structure of the SH and U stars are
very similar because both are red supergiants.

In the past, it has sometimes been assumed that the helium core mass
uniquely determines the presupernova structure of a star and hence
that structure would be roughly independent of the metallicity for two
stars that made the same helium core mass. While this is qualitatively
true, the figures here show that this assumption is not very accurate for
supernova progenitors below 30 \Msun.

\subsection{Choice of Fiducial Mass and Time for Evaluating the Compactness}
\lSect{s2.3}

With the new surveys, it is possible to address a point of possible
concern - the choice of mass (2.5 \Msun) and time (``PreSN model'')
for evaluating the compactness in \eq{compact}.  The full evolution of
a massive star to the point that its iron core collapses and possibly
powers an explosion is being characterized by a single number
here. What motivates the choice of this particular point in space and
time?

\citet{Ugl12} explored the effect of evaluating $\xi$ at different
fiducial masses and concluded that 1.75 \Msun \ might be a better
discriminant of explosion characteristics rather than the 2.5 \Msun
\ chosen here and in \cite{Oco11}. The time of core bounce rather than
initial collapse (``preSN'') also seemed a more relevant time for its
evaluation. Recently, \citet{Oco13} also examined the choice of
1.75\Msun, since the early neutrino signal is more sensitive
to the structure around the neutron star.  Obviously smaller choices
than 1.75 \Msun \ would not be sensible since they often lie {\sl
  within} the collapsing iron core itself, and depend on different
physics that occurs after that core has already reached high density,
but what about values in between?

To address these questions, the collapse of the S series was continued
until the central density reached $\rho_c = 5\times10^{11}$g
cm$^{-3}$. This is about 100 times greater than the central density of
presupernova S-series stars. Beyond this point, neutrino trapping and,
ultimately, the nuclear equation of state, would influence the
dynamics (\Fig{f7}). As expected, $\xi$ rises with
increasing central density since a smaller radius encloses the same
mass. It also rises as the mass chosen for its evaluation decreases,
since the average density enclosed by r increases faster than
r$^{-2}$. Running KEPLER well beyond $5 \times 10^{11}$ g cm$^{-3}$ to
nuclear density gives values of questionable accuracy, but strongly
suggests that very little further evolution will occur in the
compactness so long as the sampling mass remains greater than 1.75
\Msun. As the figure shows, the compactness evaluated at 2.5 \Msun
\ changes very little during the collapse from the presupernova star
to high central density. At that larger radius, the hydrodynamical
response time is longer than the time for the denser part of the core
to collapse.

\begin{figure}[htb]
\centering
\includegraphics[width=0.35\textwidth,angle=90]{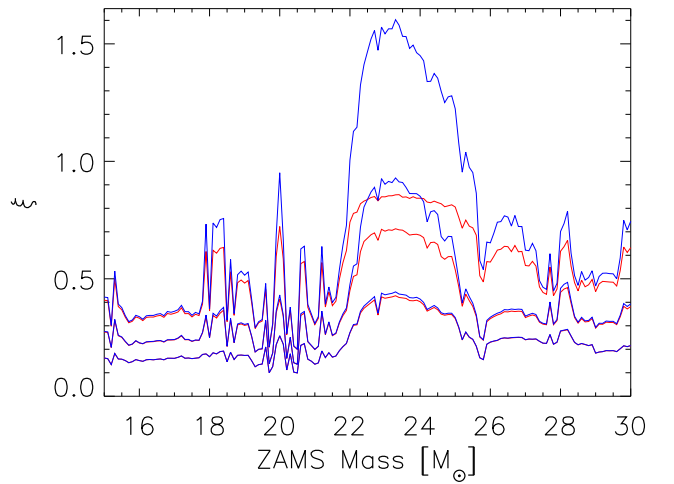}
\caption{Presupernova compactness for the S-series models evaluated
  when the infall speed reaches 1000 km$s^{-1}$ (red), and later when the
  central density reaches $5\times10^{11}$ g cm$^{-3}$ (blue). Each
  pair of curves represent $\xi$ evaluated at the same time in
  evolution, but at different points in fiducial mass: 1.75 \Msun (top),
  2.0 \Msun \ (middle) and the default 2.5 \Msun \ (bottom).  Little evolution
  occurs during the collapse from about $5.5\times10^{9}$ g cm$^{-3}$ to $5
  \times 10^{11}$ g cm$^{-3}$ if the compactness is evaluated at 2.5
  \Msun, but for smaller fiducial masses the contrasts seen at 2.5
  \Msun \ are significantly amplified with time. \lFig{f7}}
\end{figure}

Substantial variation does occur, however, for smaller sampling
masses. Smaller fiducial masses show a non-linear amplification of the
structure in the compactness curve with time. Large values get much
larger than adjacent smaller ones, suggesting the development of
islands of stars that may be hard to blow up. This is due to the
tendency of high density regions to collapse faster under the
influence of their own gravity and is particularly apparent for stars
in the 18 to 22 \Msun \ mass range. This may account for some of the
variability in outcome seen by \citet{Ugl12} for supernovae in this
mass range.

The robustness of \cp \ evaluated at 2.5 \Msun, however, and its
strong correlation with the compactness evaluated at smaller masses
suggests that we can continue to use our standard choices of time and
mass for its evaluation. It should be kept in mind, however, that
structures that seem small in some of the plots for presupernova stars
may become amplified by the further collapse.

\section{Physical Basis of the Behavior of the Compactness Parameter}
\lSect{s3}

Surveys of stellar evolution find a complex, non-monotonic behavior
for the compactness parameter as a function of main sequence mass and
metallicity. Why is this so?  Why doesn't the compactness vary
smoothly with mass as it would in a polytrope with a single index?

Four factors drive the development of the compactness profile for
massive presupernova stars. One obvious effect is the contraction of
the core to higher densities in order to burn heavier fuels. This
contraction increases the average density inside 2.5 \Msun \ and
causes \cp \ to grow with time. Another is the tendency of lighter
stars to have lower entropy cores and be more degenerate. Degeneracy
is responsible for the transition between stars that make planetary
nebulae and those that make supernovae around 8 \Msun, but the effects
of degeneracy on the post-carbon burning evolution continue to be
important throughout the entire mass range studied here. Third, as has
been noted previously \citep{Bar94,Tim96}, is the disappearance,
around, 20 \Msun, of central convective carbon core burning. Above 20
\Msun, carbon fuses away without contributing a large excess of energy
generation over what neutrinos are carrying away in the center of the
star.

\begin{figure}[htb]
\centering
\includegraphics[width=0.48\textwidth]{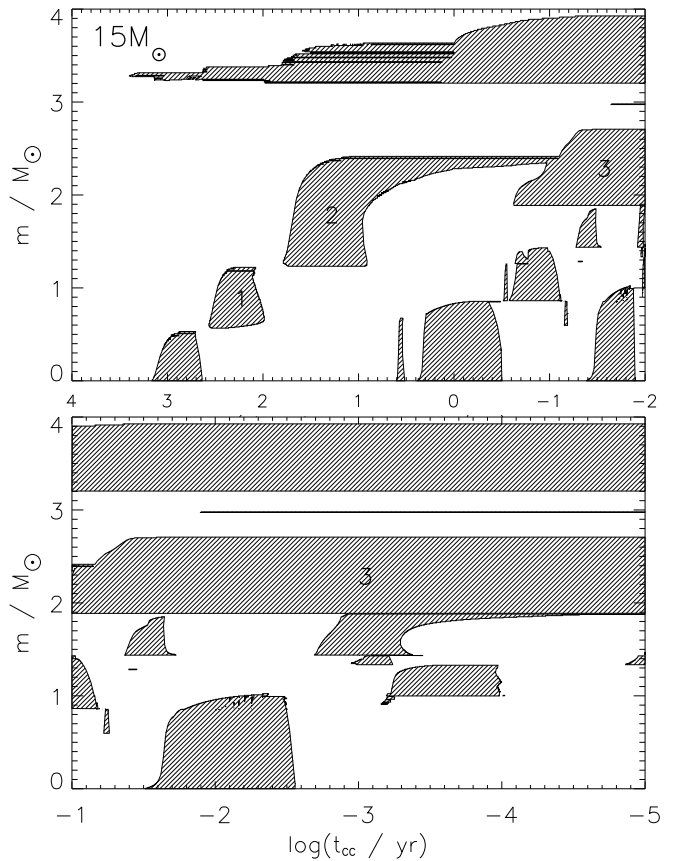}
\caption{Convective history of a 15 \Msun \ model, a typical supernova
  mass, (top:) from carbon burning until silicon shell ignition, and
  (bottom:) from the end of first oxygen shell burning until 5 minutes
  before collapse. Only the inner 4 \Msun \ is plotted and the time axis
  is plotted as the logarithm of time until iron core collapse
  ($t_{cc}$) measured in years. Cross hatched regions indicate
  convection and integer numbers annotate convective carbon burning
  shell episodes. The convective carbon core burning starts about
  1000 years before death and is followed by three episode of carbon
  shell burning. Oxygen burning starts about two years before death
  and is preceded by a brief stage of neon burning. Silicon burning is
  the last convective episode in the center. At silicon depletion, no
  oxygen shell is active, only a carbon shell at about 1.9
  \Msun. Shortly afterwards though, the third oxygen shell ignites in
  the incompletely burned ashes of the second shell. This final
  vigorous shell merges with the carbon and some neon burning shells
  shortly before the star dies.  The operation of these shells,
  especially the ones between 1.5 and 2.5 \Msun, affects the
  compactness parameter.\lFig{f8} }
\end{figure}

Fourth and frequently overlooked (though see Barkat 1994) are the
effects of convective carbon and oxygen shell burning. Lighter stars
can have three or more carbon burning shells in addition to core
carbon burning (\Fig{f8}). As these stars evolve, their central
regions become increasingly degenerate, especially after carbon
depletion in the star's center, and, by oxygen ignition, the concept
of a Chandrasekhar mass has some approximate meaning, especially for
stars lighter than 30 \Msun. Shells that burn outside the effective
Chandrasekhar mass, roughly 1.7 \Msun \ depending on thermal
corrections, but inside the point where the compactness is measured at
2.5 \Msun, will considerably modulate the compactness parameter.

\subsection{Evolution Through Central Carbon Depletion}
\lSect{s3.1}

\Fig{f9} shows that \cp \ remains nearly independent of the main
sequence mass until after carbon ignites in the core. Prior to this
time, the fiducial point at 2.5 \Msun \ lies well within a much larger
star or helium core that can be characterized by a single polytropic
index.  The tendency of $\rho_c$ to decrease as $M^{-2}$, as given by
\eq{poly}, is offset by the slight increase of the burning temperature
with mass, resulting in a nearly flat curve.

\begin{figure}[htb]
\centering
\includegraphics[width=0.35\textwidth,angle=90]{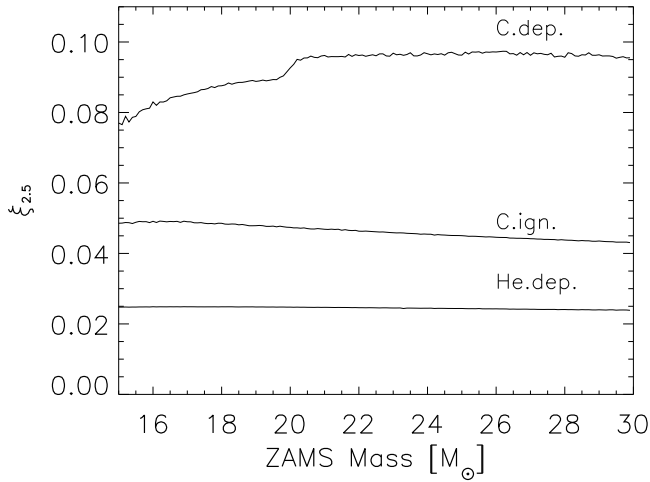}
\caption{The evolution of \cp \ as a function of main sequence mass for
  the S series models up to the point when the carbon mass fraction
  decreases below 1\% in the center of the star (``C.dep''). At each
  successive stage, the star has contracted to higher density and the
  average curve for \cp \ rises. Below about 20 \Msun \ carbon burns
  convectively in the core and this reduces \cp.\lFig{f9} }
\end{figure}

By the time carbon has disappeared from the center of the star though,
things have started to change. A pronounced dip in \cp \ develops for
the lower mass stars, and the curve shows an abrupt, small rise at 20
\Msun. For current code physics and solar metallicity, 20 \Msun \ is
the mass below which carbon burns convectively (exoergically) at the
stellar center as opposed to radiatively (endoergically). Convection
brings additional fuel into the burning region increasing the
effective supply by approximately the ratio of the convective mass to
the mass of the burning region. Because of the high temperature
sensitivity of the carbon fusion reaction, the energy generating
region is small, so the enhancement is significant. During this longer
time, neutrino losses carry away both energy and entropy, not only
from the carbon convective core, but from the hot, overlying
helium-rich layers supported by it. This loss of entropy exacerbates
the natural tendency of lighter cores to have greater degeneracy and
accelerates the development of a compact, white dwarf-like core
structure - a dense degenerate core surrounded by a much
  less dense extended envelope.

\begin{figure}[htb]
\centering
\includegraphics[width=0.48\textwidth]{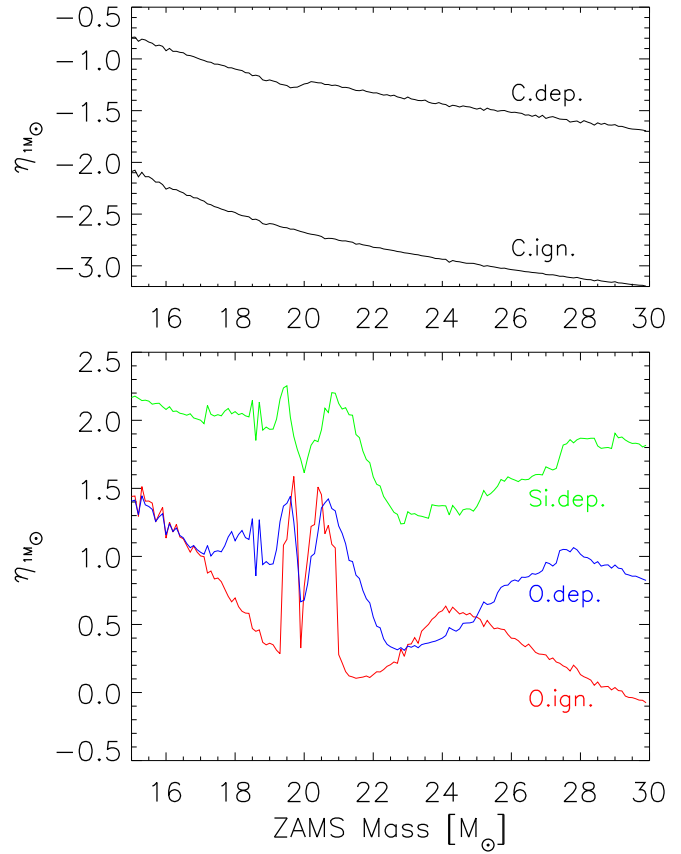}
\caption{The degeneracy parameter, $\eta\equiv\mu/kT$, averaged over
  the inner 1 \Msun \ of the star is shown as a function of initial
  mass at the times of (top:) carbon ignition and core carbon
  depletion, (bottom:) oxygen ignition, oxygen depletion and silicon
  depletion.  Very negative values of $\eta$ correspond to a
  non-degenerate gas and large positive values imply degeneracy. The
  cores of the lower mass models grow much more degenerate due to
  neutrino losses between carbon depletion and oxygen ignition. The
  structure around 20 \Msun \ in the curve for oxygen ignition reflects
  the brief appearance of strong, deeply-sited carbon convective
  shells in this mass range (\Sect{s3.2} and \Fig{f11}).  These
  degeneracy curves at oxygen ignition and depletion are nearly mirror
  images of the \cp \ curves at these times (\Fig{f13}) showing
  the key role played by degeneracy in determining the compactness.\lFig{f10}}
\end{figure}

As \Fig{f10} makes clear, {\sl central} carbon burning, by itself, is
not the whole story though. Most of the increase in central degeneracy
occurs after carbon has been exhausted in the star's center. It is
during the period between central carbon depletion and oxygen
depletion that neutrinos cool the core appreciably and, for masses
below about 30 \Msun, give it a white dwarf-like structure. During
this long cooling-off time, contraction of the inner core is
frequently held up by two, or even three vigorous carbon burning
shells (\Fig{f11}), the first igniting shortly after carbon core
depletion typically about 0.5 \Msun \ from the center. As we shall
see, the presence and location of these shells is strongly correlated
with how carbon burns in the center, so 20 \Msun \ remains a critical
mass. It is ultimately the timing and location of these carbon
convective shells, however, not just central carbon burning, that
account for the structure developed during and after oxygen burning.

\begin{figure}[htb]
\centering
\includegraphics[width=0.48\textwidth]{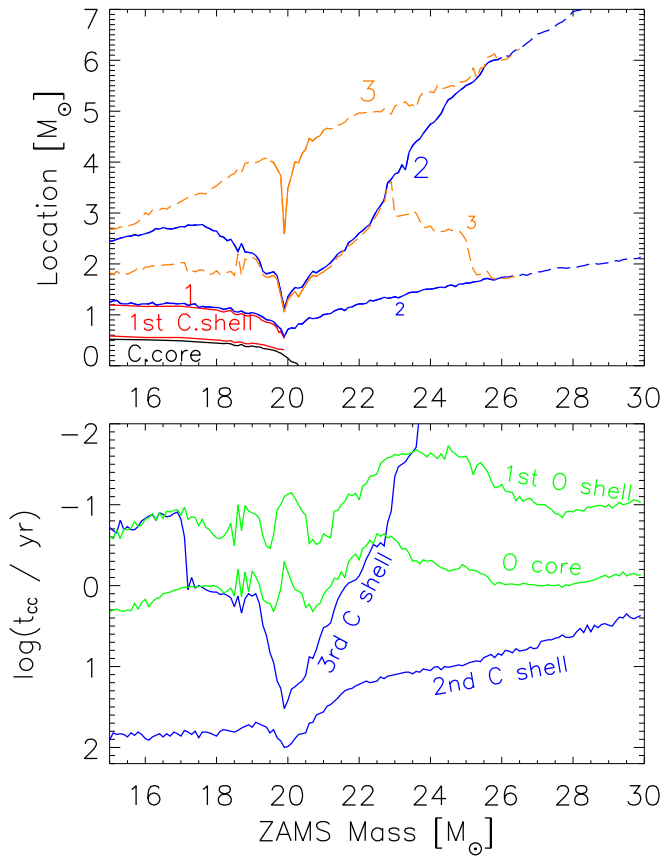}
\caption{(top:) The history of the carbon burning core and shells for
  the S-series models. Different colors show the boundaries of each
  shell: core - black, first shell - red; second shell - blue; third
  shell - orange. The outermost shell is plotted as a dashed line.
  Large numbers identify the maximum extent of shells, while smaller
  ones identify their bases. Matter between curves of the same color is
  convective at some time during the star's life when the shell is
  (even temporarily) active. The transition from convective carbon
  core burning to radiative burning near the center at around 20 \Msun
  \ greatly affects the location and extent of all three convective
  shells. For slightly larger masses, the first convective shell
  (``1'') disappears. After that mass carbon burns from the center
  radiatively out to the base of the former second shell.  The
  location of the third shell is drawn as a dashed line when it that
  shell exists for less than a year at the end of the life of the
  star.  (bottom:) difference between starting times and core collapse
  for various convective episodes (oxygen burning - green and carbon
  burning - blue) as a function of initial mass. \lFig{f11}}
\end{figure}

\subsection{Carbon Shell Burning}
\lSect{s3.2}

\Fig{f12} shows the evolution of the compactness parameter for the S
series stars from the time carbon is depleted at the center of the
star until oxygen is similarly depleted. This is a critical period
when the major features of the compactness emerge (\Fig{f7}).

\begin{figure}[htb]
\centering
\includegraphics[width=0.35\textwidth,angle=90]{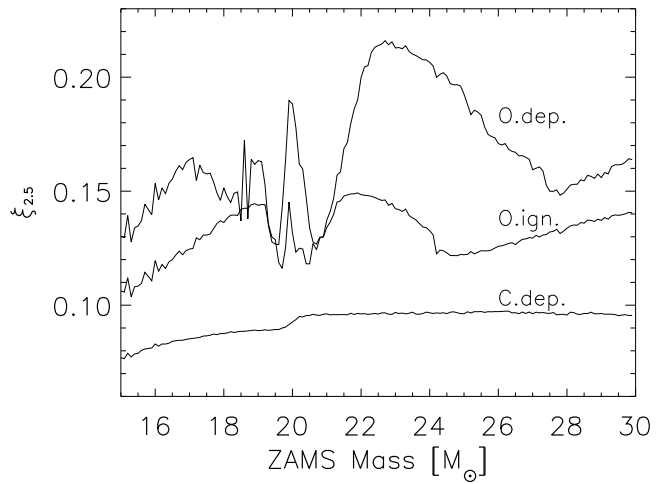}
\caption{The compactness parameter, $\cp$, as a function of initial
  mass at the time of central carbon depletion (lower curve; see also
  \Fig{f9}; oxygen ignition (central temperature equals 1.6 GK);
  and oxygen depletion (central oxygen mass fraction equals 0.04; top
  curve). Non-monotonic behavior caused by carbon burning shells is
  clearly evident above 18 \Msun \ by the time oxygen ignites, and a
  broad peak in \cp \ is emerging around 18 - 24 \Msun \ with a
  ``notch'' between 19 and 21 \Msun. Note the strong anti-correlation
  of \cp \ with degeneracy below 22 \Msun \ (\Fig{f10}) especially
  around 20 \Msun. \lFig{f12}}
\end{figure}

The typically low values for \cp \ below 18 \Msun \ result from the
operation of the first two convective carbon shells
(\Fig{f11}). While the star is supported by these shells, its
inner regions radiate neutrinos and become cool and degenerate. The
rapid variation of \cp \ in the range 18 to 21 \Msun \ reflects the
response of these three shells to the disappearance of the convective
core. When that happens, {\sl the first carbon burning shell
  disappears}, and the second and third convective shells move inwards
dramatically in response to its loss. These events do not happen
simultaneously. The center of the star ceases to be convective at 20.3
\Msun, but the first shell has already gone out at 19.9 \Msun. The
second and third shells ''migrate'' inwards until this mass
is reached, and then move outwards for bigger stars.  The offsets in
these masses and the rapid motion, first in, then out, of shells two
and three underlie some of the complex structure in degeneracy and
compactness seen around 20 \Msun.

\begin{figure}[htb]
\centering
\includegraphics[width=0.48\textwidth]{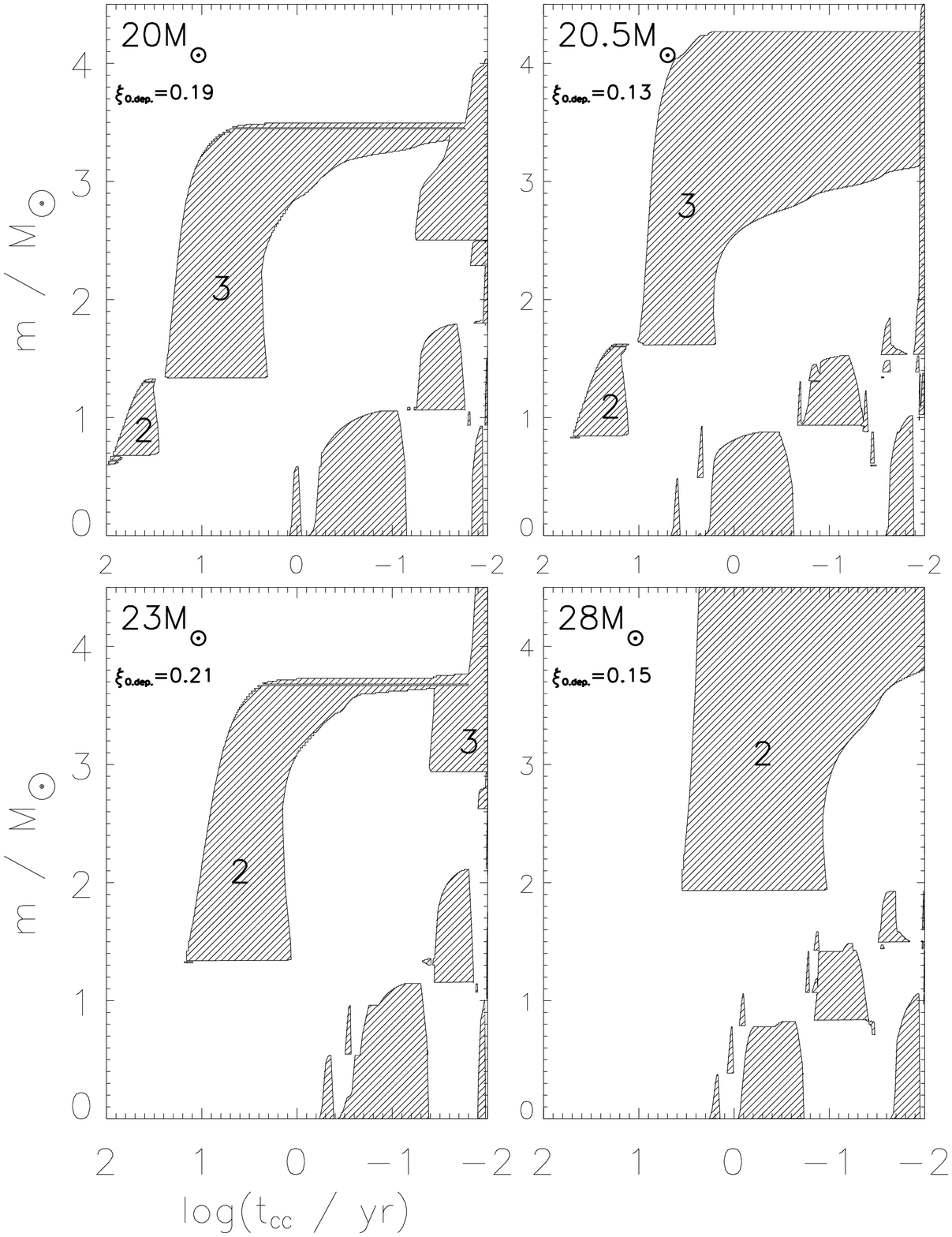}
\caption{Convective history of four models showing the major changes
  in the evolution until silicon burning. Only the inner 4.5 \Msun
  \ is plotted in each case and the format is similar to \Fig{f8}, but
  here all stars, except the 20\Msun\ model (which has a very tiny
  convective core), burn carbon radiatively in their
  centers. Prominent in all figures is a powerful carbon burning shell
  igniting roughly 10 years before the star dies. Note the migration
  outwards of the base of this shell as the mass increases. The shell
  that ignites farther out supports the star while the inner core
  emits neutrinos and becomes increasingly degenerate. The shorter
  Kelvin-Helmholtz time for the larger carbon-depleted core in the
  heavier star also leads to oxygen ignition at a substantially
  earlier time. The 20\Msun\ and 23 \Msun \ models end up with a large
  value of \cp, while 20.5\Msun\ and 28 \Msun \ models give a smaller
  value. See also \Fig{f13}. \lFig{f13}}
\end{figure}

From 21 to 30 \Msun, the compactness is most sensitive to whether,
when oxygen core burning ignites, there is, or very recently has been
an active carbon convective shell inside of about 2.0
\Msun\ (\Fig{f13}). An active shell supports the overlying star and
relieves pressure on the oxygen core, allowing it to evolve as if it
had a smaller effective mass. Smaller oxygen cores burn longer, emit
more neutrinos, cool more effectively, and become more degenerate. The
relevant carbon shell can either be the third (for 18 - 20 \Msun) or
the second (for 21 - 30 \Msun) shell.  If that shell is inside 1.4
\Msun, the oxygen core lacks sufficient mass to ignite \citep{Nomo88},
so the most important mass range for the convective shell is 1.4 to
2.0 \Msun.

The location of the boundaries of these convective carbon shells are
given in \Fig{f11}. The third convective shell is given there as a
solid curve only for the mass range where it has a lifetime longer
than one year. Shorter lived shells do not greatly affect the
evolution prior to oxygen depletion. Based upon this criterion, the
third shell only has a major effect in the range 19 to 21 \Msun. For
heavier stars, the second shell dominates (note that, for the sake of
continuity, we persist in calling it the ``second'' shell even though
the ``first'' convective shell has been replaced
  by radiative burning).

This behavior is illustrated by the convective history during carbon
and oxygen burning for stars of four masses: 20.0, 20.5, 23.0 and 28.0
(\Fig{f13}).  None of these four stars experience central
convective carbon core burning or ``first shell'' burning (in 20
\Msun\ there is a very tiny convective carbon core). The relevant
carbon shells are thus ``2'' and ``3'', and they are labeled as such
in the figure. At oxygen depletion, the 20.5 \Msun \ and 28.0\Msun
\ models have small values of \cp \ (0.13 and 0.15, respectively),
while the 20.0 and 23.0 have larger ones (0.19 and 0.21).

In the 20 \Msun \ model, shells 2 and 3 complete their major burning
about a year before oxygen ignites in the stellar center. Shell 3
is located at 1.32 \Msun, which is too small to allow oxygen to
ignite. Later, following a brief, inconsequential ``blip'' of
convective neon burning, oxygen burning ignites in a core devoid of
any nuclear energy sources out to more than 3 \Msun. The entropy is
higher and the degeneracy is less at both oxygen ignition and depletion
(\Fig{f10}). Oxygen burning ignites late, with less than a
year remaining in the star's life. Ultimately the star dies with a
relatively large compactness parameter, 0.26.

The 20.5 \Msun \ model, though only slightly different in mass, has a
very different evolution. The third carbon shell now has its base at
1.62 \Msun \ which is sufficiently large for oxygen burning to ignite
before carbon shell burning is over.  When oxygen does ignite at the
center, the resulting expansion extinguishes the carbon shell,
leaving unburned carbon in the region outside 1.62 \Msun. Some
neutrino losses have already occurred in the oxygen before it ignited, but
over the course of its burning, which commences much earlier than in
the 20.0 \Msun \ model, more cooling occurs and the core becomes
very degenerate. By oxygen depletion, the compactness parameter has a
value of 0.13, which declines still further to 0.10 when the star
dies.

For 23.0 \Msun, the evolution resembles that of the 20.0 \Msun \ model
more than it does the 20.5 \Msun \ model. By now, the third carbon
shell is igniting so late as to be unimportant. Shell 2, burning at
1.33 \Msun, however, again delays oxygen ignition to a late time. The
oxygen burns in a core that, at the time has essentially no active
burning shell except the helium shell far above it
(off-scale). It is easier for the layers above the core
  to contract since there is a time with no active burning, neither
  core nor shell. Once it ignites, the oxygen
core burns rapidly and at low density, producing a
less degenerate silicon core. By oxygen depletion, the compactness
parameter is 0.21 which rises to 0.42 in the presupernova star, the
largest compactness of any model under 30 \Msun.

For 28.0 \Msun \ and nearby masses, the evolution resembles that of
the 20.5 \Msun model, but with no third shell. The second shell has
now moved out to 1.91 \Msun \ which is enough to allow core oxygen
burning to ignite. There is always support by an active
  burning front, reducing the contraction of the outer core. Oxygen
burning lasts longer and occurs in a core of smaller effective
mass. The compactness at the end of oxygen burning is back down to
0.15, though this rises to 0.28 in the presupernova star.

Beyond 28 \Msun, the second carbon shell continues to move out and the
effective core mass when it ignites oxygen grows. When the shell
crosses 2.5 \Msun \ at 30 \Msun, the compactness parameter
rapidly increases once more. This marks the end of the epoch of
carbon shells. All heavier stars will have large compactness
parameters (\Fig{f3}) with relatively small variations introduced
by the {\sl oxygen} burning shells.

The location and strength of the carbon shells prior to oxygen
depletion thus sculpt a highly variable result out of what was a
comparatively smooth curve at carbon depletion. In some sense, 20.5
\Msun \ and 28 \Msun \ stars resemble each other more at death than do
20.5 \Msun \ and 20 \Msun \ stars. Above 30 \Msun, the carbon shells
lose their influence.

\subsection{From Oxygen Depletion to Silicon Depletion}

After oxygen finishes burning in the center, powerful convective {\sl
  oxygen}-burning shells develop that can also modify the structure
either by themselves, or by interacting with existing carbon
shells. The continued evolution serves chiefly to amplify the features
already present at oxygen ignition (\Fig{f14}). Having developed a
degenerate core, further shell burning pushes matter out at 2.5 \Msun
\ and reduces \cp. The location of the first oxygen shell at silicon
depletion correlates tightly with \cp.  Its base is located between
0.8 and 1.2 \Msun \ and is set by the extent of the oxygen convective
core. Its extent correlates tightly with the location of its
base. Similarly, the second oxygen shell, when there is one, sets atop
the extent of the first. The farther out these shells and the later
their ignition, the more extended the core, which results into a shallower 
density gradient around the iron core later on (higher \cp).
The peak in \cp \ in \Fig{f14}, around 23 to 25 \Msun, \ is
enhanced by the migration outwards in mass of this oxygen burning
shell.  Unlike the carbon burning shells, however, the outwards
movement of the oxygen burning shell is not monotonic in mass, since
the extent of the oxygen core varies due to the effect of previous
carbon shells.  As it recedes inwards from 24 to 28 \Msun, the \cp\
curve declines, leading to a pronounced dip at 28 to 30 \Msun. Above
30 \Msun, the oxygen burning shells move out rapidly.

\begin{figure}[htb]
\centering
\includegraphics[width=0.45\textwidth]{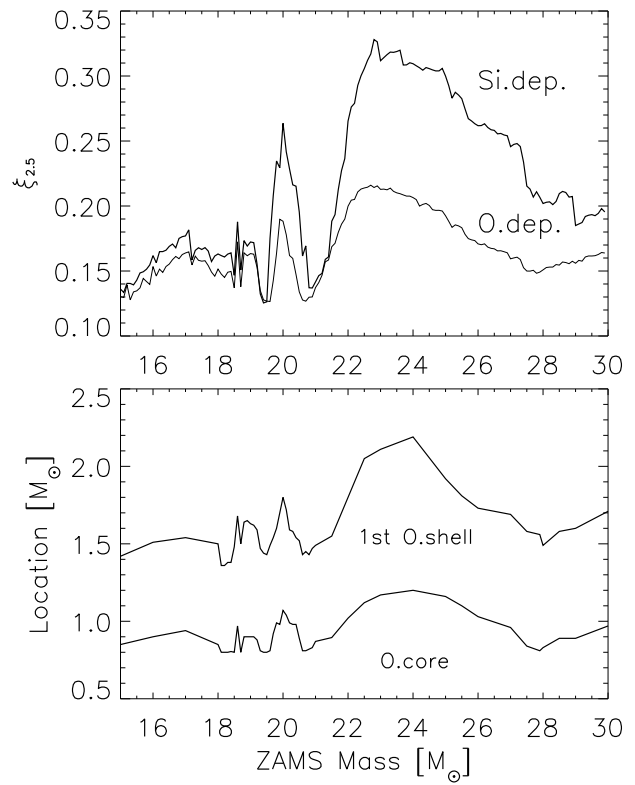}
\caption{Top: The compactness parameter as a function of initial mass
  at the times of oxygen depletion (lower curve; see also
  \Fig{f13}) and silicon depletion. The tendency of existing
  features to be amplified by the contraction, as seen previously in
  \Fig{f13}, continues. Bottom: The location of the outer extent
  of the oxygen convective core, which is also the base of the first
  oxygen shell is given as the lower curve labeled ``O.core'', while
  the extent of the first oxygen shell is the curve labeled ``1st
  O.Shell''. Note the correlation between shell locations and the
  compactness. \lFig{f14}}
\end{figure}

\clearpage

\subsection{From Silicon Depletion to Presupernova}
\lSect{s3.4}

\Fig{f15} shows that, while the peak in \cp \ at 23 \Msun \ continues
to grow due to core contraction and the operation of the oxygen
shells, the principal qualitative change in \cp \ during this final
stage of the star's life is a ``chopping up'' of the peak that existed
at silicon depletion around 20 \Msun \ into finer structures. Given
that these structures will be amplified in the collapse (\Fig{f7}),
they may yet affect the outcome of the explosion.

\begin{figure}[htb]
\centering
\includegraphics[width=0.35\textwidth,angle=90]{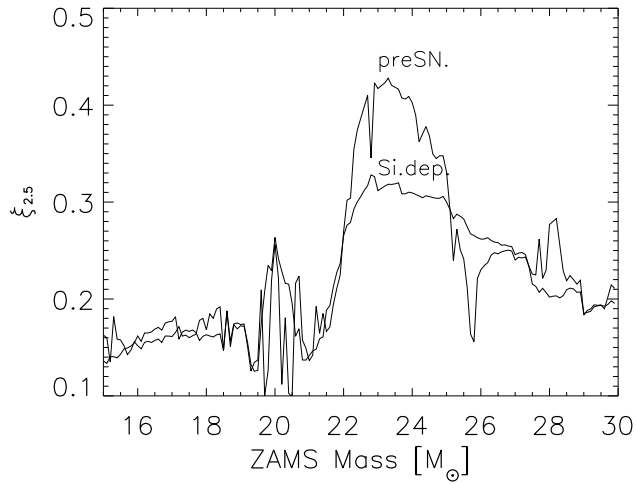}
\caption{The compactness parameter, $\cp$, as a function of initial
  mass at silicon depletion and for the presupernova star. The peak at
  24 \Msun \ grows due to core contraction and fine structure is
  imprinted between 19 and 22 \Msun \ by a strong oxygen burning
  shells in the last few days of it's life. \lFig{f15}}
\end{figure}

This rapid variation for a narrow range of masses results from an
almost random decision by the star whether to burn oxygen or silicon
in a shell first, the decision depending upon fine details of all that
happened before. \Fig{f16} shows the history of the convection
following central silicon depletion in two stars of the S series with
masses 20.1 and 20.2 \Msun. Note especially the timing of the ignition
of silicon shell burning compared with the growing extent of the
strong oxygen burning shell with a base at 1.71 \Msun \ in the 20.1
\Msun \ star and 1.59 \Msun \ in the 20.2 \Msun \ star.  In both stars
this convective oxygen shell is very powerful, and eventually grows to
include both the carbon and neon burning convective shells in one
large aggregate shell shortly before the star dies. The oxygen shell
in the 20.1 \Msun \ star drives a similar linking of convective layers
and creates a large positive energy generation rate, but only {\sl
  after } the silicon shell has already burned out and the star is
close to collapsing. In the 20.2 \Msun \ star, on the other hand, the
strong oxygen shell ignites {\sl before} silicon shell burning and
operates for a much longer time.

\begin{figure}[htb]
\centering
\includegraphics[width=0.48\textwidth]{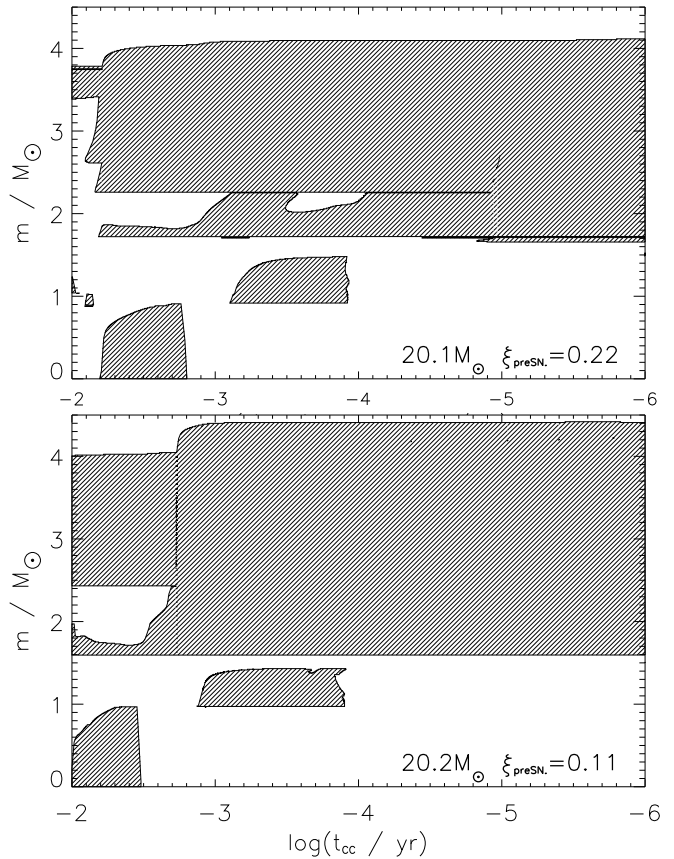}
\caption{Convection plots for the 20.1 \Msun \ (top) and 20.2 \Msun
  \ (bottom) stars of S series.  Each plot shows the evolution inside
  inner 4.5 \Msun \ during the -2 to -6 in the log of time until
  collapse. This last few days of evolution roughly corresponds to the
  time-evolution from the start of Silicon ignition in the core until
  the presupernova stage.  Shaded regions describe convection.  Notice
  in 20.2\Msun, where the \cp (measured at presupernova) is low, the
  Silicon ignites int the core earlier and the shell C-O burning is
  very different. \lFig{f16} }
\end{figure}

In the 20.1 \Msun \ star, oxygen shell burning initially generates,
only a limited convection zone that slightly alters the structure
of the star before silicon shell burning ignites and slows the
contraction. In the 20.2 \Msun \ star, the oxygen shell becomes strong
first and has already engulfed a large fraction of the CO
core before the silicon shell ignites.  The 20.1 \Msun \ star ends up
with a high \cp, the 20.2 \Msun\ star with a low one. The strong oxygen
shell in the latter pushes matter out to larger radii and leaves a
 less extended structure inside.

\begin{figure}[htb]
\centering
\includegraphics[width=0.35\textwidth,angle=90]{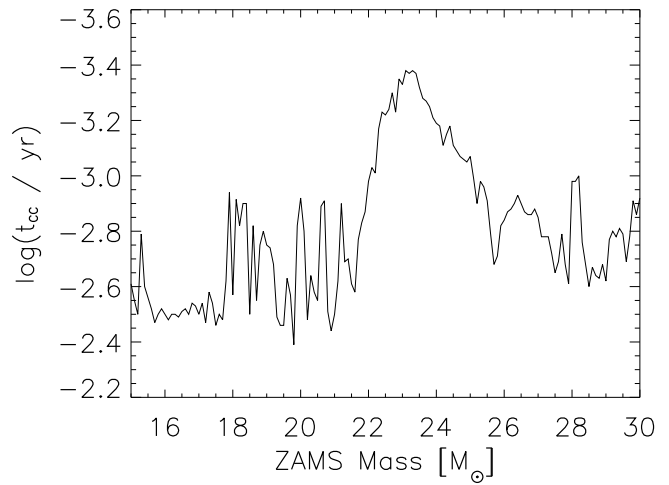}
\caption{Log of the time from the core Si depletion until the
  iron-core collapse as a function of the initial mass for S series
  models.  A striking correlation is evident with the variation of the
  compactness shown in \Fig{f2}.\lFig{f17}}
\end{figure}

This sensitivity of the presupernova compactness to the timing of
oxygen and silicon shell burning suggests that a correlation might
exist between the time from silicon core depletion to death and
\cp. This is indeed observed (\Fig{f17}). The longer the strong oxygen
shell supports the star after the initial iron core has formed from
central silicon burning, the smaller \cp \ for the presupernova star.

These energetic, merged oxygen, neon and carbon burning
  shells during the last hours of a massive star's life are a robust
  feature seen in many models of stars above about 15 - 20 \Msun, both
  here and in previous studies \citep{Woo95,Tur07,Woo07,Rau02}.  Their
  study would be an interesting, though perhaps challenging topic for
  3D simulation.

\subsection{Fine Scale Variations and Convergence}

If such small differences in the central structure of the star at
silicon core depletion can lead to major changes in presupernova
structure, even for stars differing in mass by only 0.1 \Msun, how
sensitive might the results be to other ``hidden variables'', such as
the time step criteria and zoning.  \Fig{f18} shows that, for some
masses, a substantial variation in outcome can also be caused by
varying these variables. The location of the oxygen shell is a
sensitive function of the many convective episodes that went on in the
same star during its earlier evolution, and to the physics used to
treat that convection.

\begin{figure}[htb]
\centering
\includegraphics[width=0.48\textwidth]{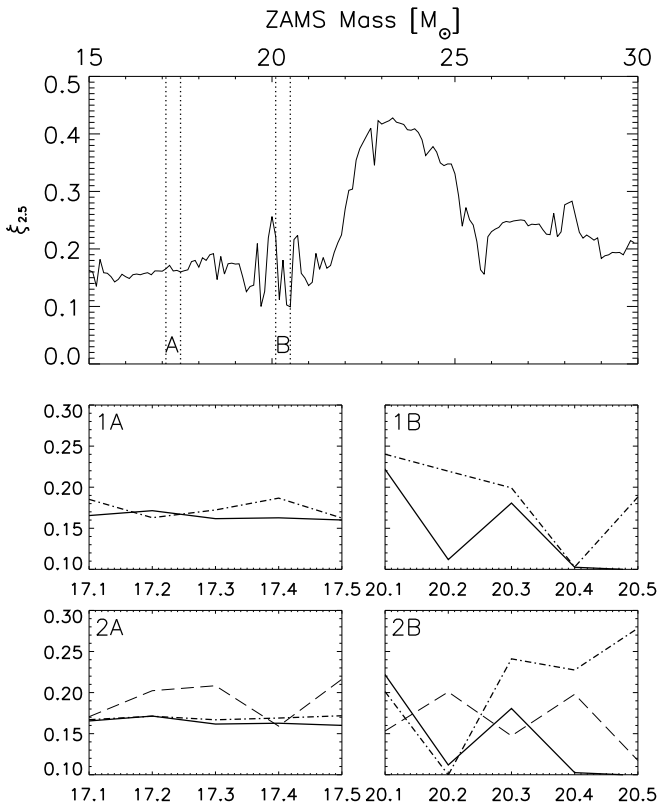}
\caption{The effect of various zoning and time step criteria on the
  final core compactness in two different regions - A: 17.1 - 17.5 \Msun \
 and B: 20.1 - 20.5 \Msun. 1A,B - different zoning: default (thick) and
  $2/3$ of default (dot-dash). 2A,B - time step: default (thick), half
  (dot-dash) and double (thick-dashed). \lFig{f18} }
\end{figure}

As previously noted by \citet{Rau02}, this sensitivity to small
changes is especially strong for the stars in the 19 to 21 \Msun
\ range. To quantify this behavior better, a subset of the S series
was calculated using the same code, compiler, and physics, but
slightly altered zoning and time step criteria.  In each case, the
standard calculations were recomputed using 50\% finer resolution
(i.e., $2/3$ the default zone size), and, in separate calculations,
the default time step was multiplied by $1/2$ and $2$.  Even these
relatively small changes in operational parameters were frequently
sufficient to provoke the star into two  different final states.
For comparison, we studied two distinct regions of main sequence
masses, 17.1-17.5 \Msun - region "A", and 20.1-20.5 \Msun - region
"B". As shown in Figure 2, the observed variations of \cp\ in these
two regions are quite different. 

Panels 1A and 2A of \Fig{f18} show that the solutions in region A are
comparatively robust. Decreasing the time step and or changing the
zoning does little to alter the final values of \cp.  Increasing 
the time step by a factor of $2$ does cause some mild variation,
$\Delta\cp<0.05$. Panels 1B and 2B of \Fig{f18}, however, show a much
more pronounced sensitivity near 20 \Msun. The resulting values of \cp\
seem to be bimodal.  Small variations in mass, zoning, or time step
can send the star down one path or the other. Typical models in the
surveys used 20,000 time steps to reach the presupernova stage and
employed approximately 1000 mass shells of variable thickness. It is
certainly feasible to double or even quadruple both the resolution and
time steps, but the values used here are already conservative. To test
the effect of finer time steps, the 21 S-series models from 19 to 21
\Msun \ were rerun with twice as many time steps. While \cp \ for some
individual stars did change significantly, the overall appearance of
the pattern, including its rapid, but bounded variation between \cp
\ = 0.1 and 0.25, did not.

\subsection{Stars Over 30 \Msun}

Above 30 \Msun, if mass loss is neglected, the structure of the core
becomes much simpler. The carbon shells have moved far out
and no longer affect the solution. The oxygen core is larger and the
oxygen shells farther out. Because of the larger mass, the degeneracy
is reduced and the compactness is always large.

The most notable feature above 30 \Msun \ is a pronounced dip in \cp
\ around 50 \Msun. The dip is not large, but is seen in both the
U-series and SH-series stars, albeit at slightly different masses
(\Fig{f3}).  It is also present in the compactness plot for the
bare CO cores studied later in \Sect{s5}.

This behavior can be traced to the presence of a strong, extended
convective oxygen burning shell during the post-silicon burning
evolution of stars over about  50 \Msun. The lighter stars lack
this shell; the heavier ones have it. Starting at 50 \Msun \ for the
U-series, this shell is present at silicon depletion with a base at
1.8 \Msun. Moving to heavier masses, the shell grows larger and its
base moves outwards, reaching 2.5 \Msun \ at 65 \Msun. There is a
sharp density decline at the base of the shell and because of this
migration outside the fiducial point for measuring \cp, the
compactness parameter rises again as the star mass passes about 60
\Msun.

Whether this shell is present or not depends upon the timing of
silicon core ignition and oxygen shell burning. Recall the key role
played by the carbon shell and oxygen ignition for stars in the range
21 - 30 \Msun \ (\Sect{s3.2}). When the carbon shell was situated
far enough out, oxygen burning would ignite before carbon shell burning
was done with major consequences for the compactness. Here, the oxygen
shell plays the role of the former carbon shell. If it burns far
enough out, the silicon core can ignite earlier. In this case,
however, igniting silicon does not blow out the oxygen
shell. It persists until the end.

\begin{figure}[htb]
\centering
\includegraphics[width=0.35\textwidth,angle=90]{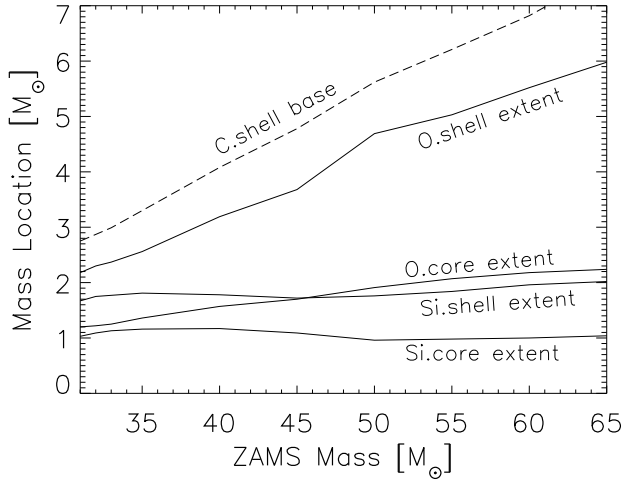}
\caption{The maximum extents of the oxygen and silicon burning
  convective cores and 1st shells are plotted against the initial
  mass of the model for SH stars.  The base of the single remaining
  carbon convective shell (dashed) is also shown, but lies well
  outside 2.5 \Msun \ and has little effect on \cp \ in this mass
  range. The bases of the 1st shells are not plotted for clarity, but
  they almost always perfectly match with the extents of the
  cores. Notice, how the silicon core size responds as the oxygen core
  overgrows the silicon shell near 45-50 \Msun. \lFig{f19}}
\end{figure}

\Fig{f19} shows the locations of various silicon and oxygen burning
episodes as a function of mass for the SH-series models. Though it
lacks the time dimension of a full convective history plot, the figure
shows that the size of the oxygen convective core increases
monotonically with mass for these heavy stars. Where the oxygen
convective shell ignites is pegged to the extent of that convective
core. At about 50 \Msun, the extent of the oxygen convective core
passes the value needed to ignite silicon core burning. An oxygen
shell ignites at the edge of that depleted core at about 1.8
\Msun. This is sufficiently large that silicon core burning also
ignites, and after that, silicon shell burning, but both ignite
without putting out the oxygen shell. Indeed, the oxygen shell
persists until the iron core collapses and forces the compactness
parameter down.

Above 50 \Msun, the compactness parameter continues to increase until
about 70 \Msun, where the onset of the pulsational pair instability
becomes important \citep{Heg10}.

\section{Sensitivity to Code Physics}
\lSect{s4}

Studies of massive stellar evolution by different groups yield
substantially different presupernova structures and nucleosynthesis. A
key quantity is the CO core mass.  The yield of
elements heavier than helium, for an energetic explosion, is the
difference between the mass of the collapsed remnant (neutron star or
black hole) and the CO core mass.  As we shall see, the compactness is
also quite sensitive to this quantity. Table 2 shows the disparate
values for the CO core mass obtained by various groups studying the
problem.  For example, a 20 \Msun \ star may end up with a CO core of
anywhere from 2.71 to 6.59 \Msun \ for various choices of convective
physics and assumptions about rotation. Even for non-rotating stars
the range is 2.71 to 4.99 \Msun. Why are the values so different?
Might using the CO core mass as a discriminant for the presupernova
compactness alleviate some of the sensitivity to uncertain code
physics?

The effect of rotation is not surprising. Rotation adds new mixing
processes \citep[e.g.][]{Heg00} that increase the masses of both the
helium and CO core. Shear mixing in differentially rotating stars can
also erase some of the sensitivity to the treatment of semiconvection
and convective shell boundaries. The large residual range of CO core
masses in non-rotating stars reflects these uncertainties in the
treatment of convection and warrants discussion.

\subsection{Sensitivity to Semiconvection and Overshoot Mixing}

Uncertainties in the treatment of convection for non-rotating massive
stars can be lumped into three categories: a) the use of
time-dependent mixing length theory (MLT) to describe the overall
transport; b) uncertainties in semiconvection, i.e., what to do about
stellar regions that are unstable to convection by the Schwarzschild
criterion, but not the Ledoux criterion; and c) the treatment of
convective boundaries (i.e., overshoot and undershoot). A detailed
review is beyond the scope of this paper. Most modern studies use
time-dependent MLT, if only for the lack of a viable, implemented
alternative. Semiconvection and convective overshoot can sometimes
have similar effects since both allow convective regions on their tops
and bottoms to grow smoothly when circumstances warrant it, and
prevent the unrealistic fracturing of convective regions into multiple
concentric shells.  Semiconvection additionally slowly mixes regions
that, owing to composition barriers, would not have mixed if the
Ledoux criterion were strictly applied.

A relevant case in point is the convective helium core. It has been
known for some time that the CO core mass is quite sensitive to the
efficiency of semiconvection during helium burning
\citep{Langer89}. In situations where semiconvection is very small (or
zero), a mathematical instability can develop that leads to the
bifurcation of the helium convective core and the production of an
unusually small CO core. In an algorithm that looks at the composition
to determine convective instability (i.e., Ledoux criterion), a single
zone that is marginally unstable might temporarily become flagged as
stable due to a trivially small increase in the mean atomic weight
(helium plus carbon and oxygen) in the zone beneath. If this zone is
flagged as stable against convection, burning during the next time
step will increase the atomic weight farther in the region that is
still convectively coupled to the burning at the center of the star
without raising it in the outer core. This makes convection between
the two shells even more difficult. The helium convective core splits
permanently into two pieces. In the outer one, helium never burns to
completion and the resulting CO core for the star is small. If enough
semiconvective mixing (or rotation) is included though, the splitting
of the the convective core does not happen. In fact, considering the
real situation in three dimensions, this numerical instability seems
unphysical and should be avoided.

Overshoot mixing can also help bridge this artificial segregation. It
also leads to an increased growth of the helium convective core at the
end of helium burning that affects not only the CO core mass, but the
carbon mass fraction when carbon burning ignites.  Simulations by
Meakin and collaborators \citep[e.g.][]{Meakin07,Via13} confirm that some
type of mixing will always take place at convective boundaries defined by
either the Ledoux or Schwarzschild criterion, although the exact
formulation of this mixing remains uncertain.

The treatment of convective overshoot and semiconvection in KEPLER has
been discussed previously \citep{Wea78,Woo88,Woo02}.  Semiconvection is
parametrized by a diffusion coefficient that is manufactured from the
local convective diffusion coefficient and the radiative one:
\begin{equation}
D_{SC} =  q_r D_r D_c /(D_c + q_r D_r),
\lEq{semid}
\end{equation}
where $D_C$ is the convective diffusion coefficient the zone would
have had based upon the Schwarzschild criterion, and $q_r$ is a
free parameter that multiplies the radiative diffusion coefficient,
$D_r$.  Usually $D_c \gg D_r$ so that $D_r$ dominates the
transport. The actual value of $q_r$ is not known, or even that such a
simple formula captures the essential features of semiconvection. In
past surveys using KEPLER, $q_r$ has usually been taken to be 0.1,
and, unless otherwise specified, that is the value used in this paper.
In practice, this generally assigns the semiconvective diffusion
coefficient a value of near 10\% of the radiative one.

Overshoot mixing is presently treated very crudely in KEPLER.  Single
zones at the top and bottom of regions flagged as unstable to Ledoux
convection are slowly mixed with a diffusion coefficient given by
\eq{semid}. The convective diffusion coefficient is calculated based
upon the Schwarzschild criterion:
\begin{equation}
w = \frac{dp}{P} - \frac{\Gamma_2}{1 - \Gamma_2} \frac{dT}{T}
\end{equation}
with $w$ assumed to be a parametrized fraction,
\begin{equation}
w = q_{Ov}  d ln(T),
\end{equation}
and with $q_{Ov}$ having a default of 0.01. Typically the convective
diffusion coefficient so calculated is much greater than $q_r D_r$ so
that the mixing occurs on a time scale given by this quantity.
This prescription lacks a physical basis, but at least allows
convective zones the liberty of slowly growing into regions where
mild entropy gradients would otherwise disallow mixing. 

To explore the sensitivity of the compactness and the CO core mass to
semiconvection and overshoot mixing, a subset of 16 solar metallicity
models in the mass range of 15 to 30 \Msun \ was calculated multiple
times using various settings for the both semiconvection and
convective overshoot mixing multipliers. The masses studied were the
integers from 15 to 30 and a total of 176 new models was calculated
(Tables 1 and 2). In these, the semiconvective multiplier, $q_r$, was
varied from 0.001 to 0.1 (16 models each with $q_r$ = 0.001, 0.0025,
0.005, 0.0075, 0.01, 0.0125, 0.015, 0.0175, 0.02, 0.05, and 0.075 in
addition to the standard models with $q_r$ = 0.1) and are collectively
called SS series.  Additionally, a set of 16 models was run with
standard semiconvection ($q_r$ = 0.1), but zero overshoot mixing
($q_{\rm Ov}$) = 0 (SO series).  All models were generated on the zero
age main sequence and run until core collapse. The CO core mass does
not change significantly after carbon depletion, however, and
presupernova values can safely be compared with other studies that
stop at an earlier time.

\Fig{f20} shows that the compactness exhibits great deal of
diversity depending on the efficiency of semiconvection.  The mass
resolution in these figures is significantly degraded by considering
only 16 masses points between 15 and 30 \Msun, but the location of the
peak in \cp, when one exists, is clear. Reduced values of $q_r$ give
smaller CO cores, more than spanning the space of published values.
The carbon mass fraction at carbon ignition is also affected by the
choice of convection physics, both in response to the altered CO core
mass and the mixing of helium at the outer boundary of the helium core
that is altered when semiconvection is turned off.  Consequently the
compactness, which is quite sensitive to both the CO core mass and the
carbon mass fraction, vary significantly.

\begin{figure}[htb]
\centering
\includegraphics[width=0.48\textwidth]{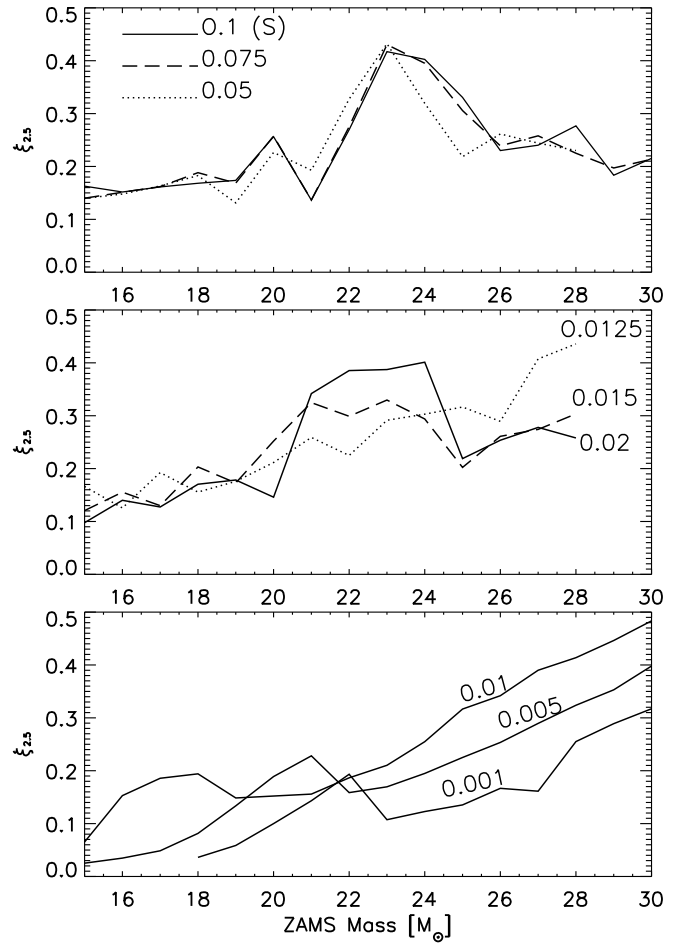}
\caption{The core compactness curve as a function of initial mass
  sensitively depends on the semiconvection efficiency. Several sets
  of SS series KEPLER models of low semiconvective efficiency (down to
  $q_r=0.001$) are shown in comparison with the S-series models (top
  panel) with $q_r=0.1$. The S-series is plotted with the
  same $1\Msun$ increments as was used in SS series
  calculations. \lFig{f20}}
\end{figure}

For small reductions in $q_r$, the peaks found for \cp \ in the
standard runs is robust. Both the locations and amplitudes do not vary
much. Though not explored, we expect that small changes upwards of
$q_r$ would have given similar results. When $q_r$ declines below 0.02
however, significant changes occur.  With smaller $q_r$, it takes a
higher mass main sequence star to give a specific CO core mass, so the
peak that was at 23 \Msun \ shifts rapidly to higher values,
ultimately leaving the grid. By $q_r$ = 0.01, the peak is no longer
clearly discernible in the mass range examined.

Continued reduction in $q_r$ (bottom panel, \Fig{f20}) results in a
compactness curve that loses most of its non-monotonic structure and
increases steadily with mass. However, such small values imply a CO
core that is small compared with all previous studies in the
literature (Table 2). For $q_r$ = 0.001, the 15, 16 and 17 \Msun
\ models closely resembled previous models calculated with KEPLER and
other codes for stars with mass near 10 \Msun
\ \citep{Nom87,Poe08,Wan09}. Oxygen ignition occurs off center in
these models and propagates as a convectively bounded flame to the
stellar center \citep[e.g.][]{Tim94}. A similar result was obtained
for $q_r$ = 0.0025 for the 15 and 16 \Msun \ models.  Such stars are
known to have a very steep density gradient surrounding the iron core,
which might be a boon to those trying to blow up stars in a wider mass
range, but contribute almost nothing to the nucleosynthesis of
intermediate mass elements. Were the observational limit on the
maximum mass supernova to be pegged not far above 18 \Msun
\ \citep{Bro13}, models of this sort would be inconsistent with the
theory of stellar nucleosynthesis. A model with 12 \Msun \ and $q_r$ =
0.001 failed even to ignite oxygen burning and would likely become an
AGB star with a degenerate neon-oxygen core and not a
supernova. Since this would be in contradiction with a large volume of
observational data showing that stars down to about 8 \Msun \ do
explode, we must consider values of $q_r$ much below 0.01 as
unrealistic.

Turning off convective overshoot mixing in KEPLER also has a
significant effect on the CO core mass and carbon mass fraction,
though not so extreme as that caused by large variations in
semiconvection (Table 2).  \Fig{f21} shows the CO core sizes and
final presupernova \cp\ of the SO-series models compared with those
from S series. The elimination of overshoot reduces the CO core sizes
(top-panel), and as a consequence, the \cp \ curve is shifted to
higher masses. For example, central carbon burning shifts to being
radiative at a higher mass compared to the S-series models with
overshooting.

\begin{figure}[htb]
\centering
\includegraphics[width=0.48\textwidth]{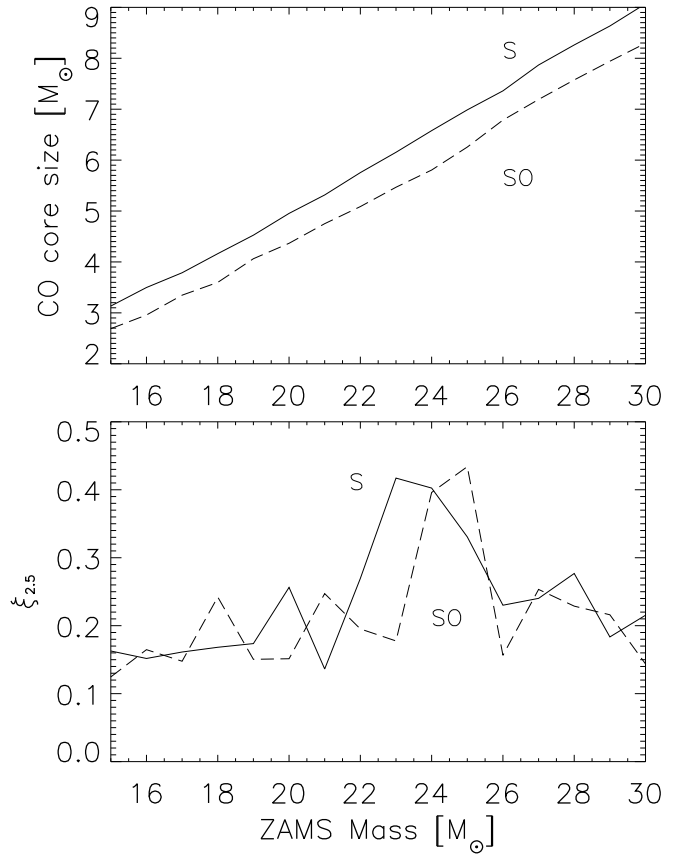}
\caption{The effects of overshoot mixing on the S-series models. The
  ``SO-series'' of models includes no overshoot mixing. (top:) CO core
  mass as a function of main sequence mass. Without overshoot mixing,
  helium burning extends to a smaller mass when burning convectively
  in the center of the star resulting in CO core masses that are
  systematically smaller. (bottom:) The smaller CO cores for a given
  main sequence mass result in a shift upwards for the compactness
  parameter plot of 1 to 2 \Msun. \lFig{f21}}
\end{figure}

These results, which imply that the success or failure of exploding a
presupernova model of given mass will vary appreciably depending upon
who calculated the model and the description they used for convection
physics, are troubling. However, semiconvection and overshoot mixing
are real physical processes that will eventually be better
understood. Some of the very small values for $q_r$ used here may not
be physically realistic \citep[e.g.][]{Bie01} and the high threshold
mass that they imply for making a supernova is inconsistent with
observations.  Zeroing overshoot mixing is also
unrealistic. Bifurcation of the helium convective core probably does
not occur. Real stars rotate, and rotationally induced mixing reduces
the sensitivity to semiconvection and overshoot mixing. We believe
that our standard choices for $q_r$ and $q_{\rm Ov}$ (0.1 and 0.01)
are a good compromise for non-rotating stars since they avoid the
unphysical splitting of the convective core, agree reasonably well
with the results for rotating models (Table 2), and are not presently
at odds with any existing physical prediction. Clearly this is an area
where more work is needed. See \citet{Zau13} and references therein
for some recent insights.

\begin{figure}[htb]
\centering
\includegraphics[width=0.48\textwidth]{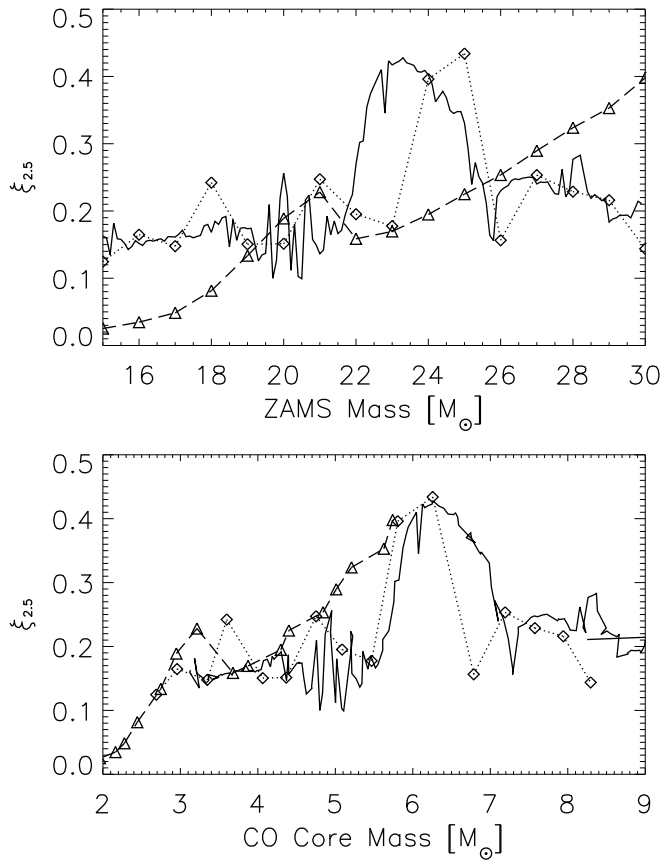}
\caption{Final \cp\ as a function of initial main sequence mass (top)
  and final CO core mass (bottom) for the S-series models
  (continuous), a subset of SS-series models with $q_r=0.005$
  (triangles) and SO-series models (diamonds).  Though the definition
  of boundary of the CO core inside a massive star is ambiguous, its
  mass can be a better discriminant of \cp \ than the initial mass in
  most cases.  \lFig{f22}}
\end{figure}

Our results also suggest that \cp, and presupernova structure in
general, will be more similar for two models having the same helium
and CO core masses than for two that simply started with the same main
sequence mass. Carrying this to the extreme, the outcome of the
explosion of two stars with the same CO core mass will be more similar
than for those having the same main sequence mass, but different CO
core masses. \Fig{f22} shows this to be approximately true.
The offset in peaks seen in \Fig{f21} for the stars calculated with
and without overshoot mixing is greatly reduced if \cp is plotted
against CO core mass rather than main sequence mass. Similarly, the
large discrepancy with the S series found for low values of
semiconvection (e.g., $q_r = 0.005$; \Fig{f20}) is significantly
improved. Note that the allowed range of CO masses in the figure is
much more limited for 15 to 30 \Msun \ stars if semiconvection is
reduced significantly. Just what constitutes the ``CO core mass''
inside a massive star is ambiguous however, because there usually is
no sharp decline of pressure at its edge (\Sect{s5}). Here we arbitrarily took 
the value inside of which the carbon mass fraction rose above 0.20.  Differing 
values of the carbon mass fraction at carbon ignition due to changes in the 
initial composition or code physics must also be considered.

\subsection{Survey of Solar Metallicity Stars Using MESA}

Even similar physics often has different implementations in various
stellar evolution codes, so to further explore the sensitivity of the
compactness curve to the way it is calculated, we repeated part of our
study using a completely different code.  MESA \citep{Pax11} is a
popular code for the study of stellar evolution which has been
successfully applied to a large variety problems in astrophysics and
is freely available. MESA has recently been applied to the calculation
of presupernova models \citep{Pax13,Ibe13,J13}. It uses similar
nuclear physics to KEPLER for energy generation during hydrogen
through oxygen burning and a very similar equation of state and
opacity tables. It can be constrained to use similar mass loss
rates. Here the red giant mass loss rates of \citet{NdJ90} was used in
both codes:
\begin{equation}
log(\dot{M})=-7.93+1.64log(\frac{L}{L_{\odot}})+0.16
log(\frac{M}{\Msun})-1.61 log(T_{eff})
\end{equation}

However, the two codes have quite different prescriptions for
semiconvection and overshooting. Given the previous discussion,
those differences can be expected to cause changes that might be
educational. MESA also uses different nuclear physics during
silicon burning and iron-core formation that affect the post-oxygen
evolution. In particular, MESA continues to use a 19
isotope network to describe the energy generation and changes in
composition after oxygen depletion, while KEPLER uses a 125 isotope
``quasi-statistical equilibrium'' (QSE) network \citep{Wea78}. The
latter is more stable and less prone to spurious temperature
transients and is also able to follow accurately the appreciable
changes in electron mole number that happen in these late phases.

Since the major features of the pre-supernova \cp(M) \ curve,
including the existence and location of peaks, have already been
imprinted by oxygen depletion, we elected to compare the results of
the two codes at that point, not for the pre-supernova star. This way,
the post-oxygen burning differences are avoided and also the run time
greatly reduced.  MESA is a continually evolving code while KEPLER is
more ``mature'', and for this study we used MESA version 4930. The
stars studied are a subset (Table 1) of our S series, that is,
non-rotating stars with solar metallicity.

This version of MESA employed Ledoux convection. For the
  mixing length parameter, $\Lambda\equiv\alpha_{MLT}H_P$, the free
  parameter $\alpha_{MLT}$ was taken to be 2.0, which is within the
  range found in the literature. In those zones flagged as
semiconvective, the \citet{Langer83} model was used. This calculates
the diffusion coefficient as:

\begin{equation}
D_{SC}=\alpha_{SC}\left(\frac{2acT^3}{9\kappa\rho^2
  C_P}\right)\frac{\nabla_T-\nabla_{ad}}{\nabla_L-\nabla_T},
\end{equation}

where $C_P$ is the specific heat at constant pressure and the
efficiency is tuned by the free parameter $\alpha_{SC}$.  We carried
out a brief survey varying this parameter between $10^{-5}$-$10^2$,
and found its effect to be rather complicated. For the masses tested,
this range of $\alpha_{SC}$ gave a spread of about $0.5\Msun$ in the
CO core mass, $1\Msun$ difference in final mass, and a one order of
magnitude change in radius. In general, for larger values ($\geq0.5$)
the MESA results approached that of the pure Schwarzschild case. The
very low values, on the other hand, gave results that were very
similar to one another, but quite different results were obtained
using the larger values. The greatest sensitivity for the CO core size
was found to be between $0.01$ and $0.5$. In our survey $\alpha_{SC}$
was taken to be $0.1$ as a ''midpoint'' between these two extremes,
which is also well within the range used in literature
\citep[e.g.][]{Yoon06}. We also note that the work to eliminate or
greatly constrain this parameter is underway \citep{Wood13,Spruit13}.

Overshoot mixing in MESA is accomplished either by fully mixing the
zones flagged as overshooting (''step'' overshooting) or by applying
an exponential cutoff to the diffusion coefficient calculated for
convection in zones that are unstable by the Ledoux or Schwarzschild
criterion. In our survey we have employed the exponential decay
formalism, that is:

\begin{equation}
D=D_{conv.}exp(-2z/fH_P),
\end{equation}

where $D_{conv.}$ is the diffusion coefficient from MLT, $z$ is the
height, and $H_P$, the pressure scale height. The efficiency is then
controlled by the free parameter $f$
(see \citet{Herwig00},\citet{Pax11} for further details).

\begin{figure}[htb]
\centering
\includegraphics[width=0.48\textwidth]{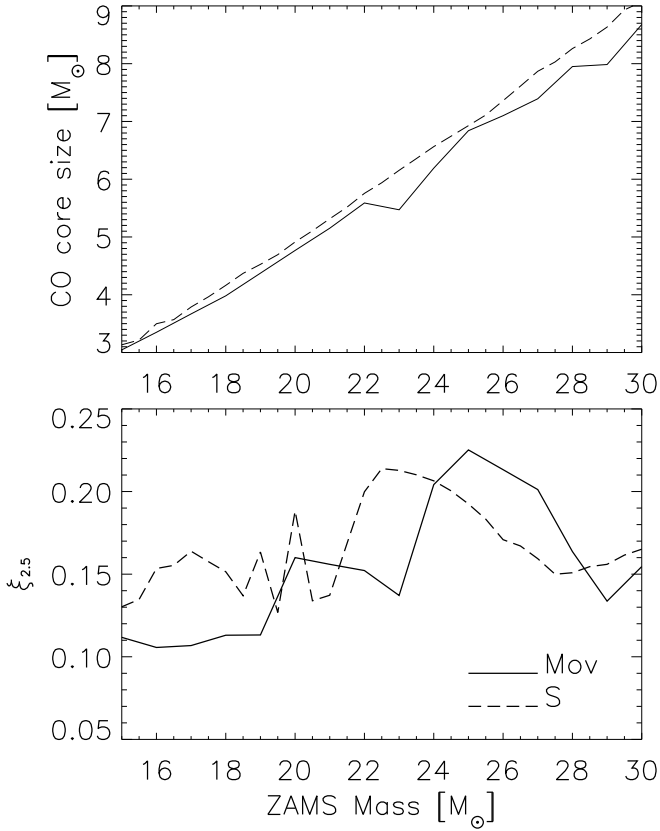}
\caption{The MESA Mov series (with overshooting) is shown in
  comparison with the KEPLER S series. (top:) CO core mass as a
  function of main sequence mass. The overshooting configuration in
  Mov produces quite similar CO core sizes as the S series
  (bottom:) The compactness parameter, \cp, as a function of initial
  mass curves at oxygen depletion in the core for each
  series. \lFig{f23}}
\end{figure}

In general, the value of $f$ is quite uncertain as well
\citep{Meakin11}.  A value of 0.016 was used by \citet{Herwig00} for
his studies of AGB stars. Since the stellar context here is quite
different, we have performed some additional test runs.  The effects
of five different choices for $f$ ranging from 0 to 0.025 were
explored (Table 2). In the end, a value of 0.025 was selected for
detailed study since it produced very similar CO core sizes to the
S-series models calculated using KEPLER(\Fig{f23}). A value as small
as Herwig's had little effect and values much bigger than 0.025
resulted in the bump in compactness around 23 \Msun \ becoming very
broadened and shifted to lower masses. Though our understanding of
overshoot mixing is not adequate to rule this out, there was no
compelling region to use larger values that would have given larger CO
core masses for the heavier stars than previously published by others
(Table 2).

\begin{figure}[htb]
\centering
\includegraphics[width=0.35\textwidth,angle=90]{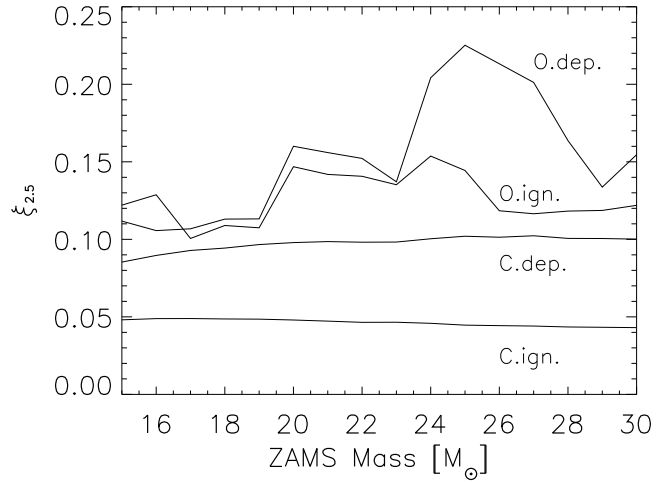}
\caption{Evolution of the compactness parameter in the MESA models of
  Mov series. Shown is the compactness parameter, \cp, evaluated at
  carbon ignition, carbon depletion, oxygen ignition and oxygen
  depletion. Compare with \Fig{f9} and \Fig{f13} for the
  equivalent runs using KEPLER. \lFig{f24}}
\end{figure}

Two sets of MESA models were ultimately calculated - one
with overshooting (Mov series) and another without (M series). Each
set had 16 models in the mass range of $15-30 \Msun$ with $1 \Msun$
increments, and both used the Ledoux criterion (Table 1). As expected,
the results from KEPLER and MESA were nearly identical during hydrogen
and helium burning, but started to diverge following helium depletion
when the structure became more sensitive to the treatment of
semiconvection and overshoot mixing.

As shown in \Fig{f24}, by carbon depletion, the discrepancy had become
significant, especially around the critical mass for the extinction of
central carbon convection, 20 \Msun, and continued to grow during
oxygen burning.

\Fig{f23} and Table 2 show the comparison at oxygen depletion
between the KEPLER S series and MESA Mov series. The quantity \cp
\ for the Mov series shows a non-monotonic structure similar to that
in the S series, but shifted in mass by about 2\Msun. Since the CO
cores are nearly identical for Mov- and S-series models, the difference
must result from either a different composition at carbon ignition or
a different treatment of convection after carbon ignites, or both.

The carbon mass fractions at carbon ignition are indeed
different in the two studies. Mass fractions in the Mov series range
from 0.34 to 0.24, declining roughly linearly as the mass rises from 15
to 30 \Msun, while the corresponding limits for the S series are 0.21
to 0.18. The larger carbon abundance probably accounts for most of the
shift in the peak of the Mov plot.  As we shall see (\Sect{s5}),
however, the treatment of post-carbon burning convection also
matters. Even bare CO cores with identical initial compositions often
have significantly different compactness when evolved to oxygen
depletion in the two codes.  The results from the M series are not
shown in \Fig{f23}, but Table 2 shows that the resulting CO cores
are much smaller, and the compactness thus differs appreciably,
especially for the lighter stars.

Despite the shift, it is encouraging that the non-monotonic structure
of \cp \ as a function of mass in \Fig{f23} and, qualitatively,
the location of its peak, can be achieved using standard settings for
convection physics in two very different codes. The study also
highlights the dependence of the outcome on the CO core mass. 

\subsection{Sensitivity to Uncertain Nuclear Physics - 
$^{12}$C($\alpha,\gamma)^{16}$O}

Because of the sensitivity of the results to the carbon mass fraction
at carbon ignition, outcomes will depend on the rates used for its
creation and destruction.  Three major reaction rates are involved,
each of which has some associated uncertainty: 3$\alpha$,
$^{12}$C($\alpha,\gamma)^{16}$O, and $^{12}$C+$^{12}$C.

It is the competition of 3$\alpha$ and $^{12}$C($\alpha,\gamma)^{16}$O
that sets the ratio of carbon to oxygen produced by helium
burning. Increasing the former raises the carbon yield while
increasing the latter makes it smaller. As we have shown, small reductions
in carbon mass fraction tends to shift the compactness curve
downwards, especially the critical mass, 20 \Msun, above which carbon
burns radiatively.

The effects of varying the three uncertain reaction rates on massive
stellar evolution, and nucleosynthesis in particular, have been
previously explored by \citet{Wea93}, \citet{Tur07}, and \citet{Wes13}
for $3 \alpha$, and $^{12}$C($\alpha,\gamma)^{16}$O and by
\citet{Pig13,Ben12} for $^{12}$C+$^{12}$C, but none of these works
explicitly focused on how these rates affect the {\sl structure} of
presupernova stars. We will not attempt a survey of all possibilities
at the present time, but focus here on just the one rate for
$^{12}$C($\alpha,\gamma)^{16}$O.

\begin{figure}[htb] 
\centering
\includegraphics[width=0.35\textwidth,angle=90]{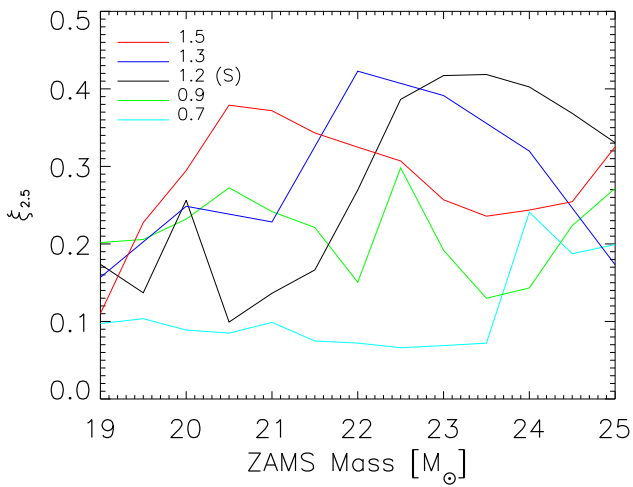} 
\caption{The compactness parameter as a function of main sequence mass
  for non-rotating, solar metallicity stars using variable rates for
  the $^{12}C(\alpha,\gamma)^{16}O$ reaction rate. Curves are labeled
  by a multiplier times the rate of \citet{Buc96,Buc97}. A value of
  1.2 was used in the rest of the paper. The current 1$\sigma$ range
  is approximately 0.9 to 1.3. \lFig{f25}}
\end{figure}

Historically, the large possible range for this reaction rate has been
a source, perhaps even {\sl the} major source of nuclear physics
uncertainty in massive stellar evolution. Recent developments
\citep[e.g.][]{Sch12} have narrowed the error bar for this rate to
less than 20\%, which is not so much greater than the uncertainty in
3$\alpha$ \citep{Wes13}. For small errors in the rates, it is the
ratio of these rates that matters most, so variation of
$^{12}$C($\alpha,\gamma)^{16}$O can act, qualitatively, as a surrogate
for exploring the uncertainty in 3$\alpha$.

The standard value used in this paper for
$^{12}$C($\alpha,\gamma)^{16}$O is 1.2 times \citet{Buc96,Buc97} who
used S(300 keV) = 146 keV b; that is, our effective S-factor at 300
keV, near the Gamow energy for helium burning, is 175 keV b.
\citet{Sch12} now report laboratory measurements and analysis that
gives S(300 keV) = 161 $\pm$ 19 stat - 2 sys + 8 keV b, or a range 0.9 to
1.3 times the Buchmann value.  A subset of our S-series models was
thus recalculated with varying multipliers on the Buchmann
rate. Thirteen different masses of presupernova star were calculated
in the mass range of 19 to 25\Msun\ (0.5\Msun\ intervals) for each
multipliers of 0.7, 0.9, 1.3, 1.5 and 1.9 times the Buchmann rate (See
Table 1). These are the SB series (Table 1).

\Fig{f25} and Tables 2 and 3 show the results. Decreasing the
rate for $^{12}C(\alpha,\gamma)^{16}O$ significantly increases the
carbon mass fraction at helium ignition, but affects the mass of
the CO core very little.  The greater (or smaller) carbon mass
fraction shifts the peak in \cp, formerly seen at 23 \Msun \ for a
multiplier of 1.2 to higher (or lower) values. For a multiplier of
1.9, the peak would disappear altogether, and is not plotted for
clarity.  However, such large values are outside the current
experimental range. The current (1 $\sigma$) error bar for the rate
translates into a multiplier between 0.9 and 1.3, but even this
smaller range can shift the explodabilty of massive stars enough
to have a dramatic potentially observable effect on presupernova
masses and nucleosynthesis. A total uncertainty of less than 10\% in
the combined error of 3$\alpha$ and $^{12}C(\alpha,\gamma)^{16}O$ may
be needed to pin down the mass range of supernovae that explode by the
neutrino transport model to an accuracy of less than 2 \Msun.

\section{Surveys with Bare Carbon-Oxygen Cores}
\lSect{s5}

The importance of the CO core mass motivates a separate study of the
compactness of presupernova stars evolved from {\sl bare} CO stars of
constant mass. It is to be emphasized that there is no direct
correspondence between the results of evolving an isolated CO core and
those obtained for a complete massive star with the same CO core mass
embedded inside a helium core. The lack of a precipitous drop in
density and pressure at the edge of a CO core inside a massive star
results in an evolution that is qualitatively different. The
definition of a CO core there, even by composition, is ambiguous
\citep{Hirschi04}. How does one count a partly burned helium shell
(and these are a common case)?  Nevertheless, CO cores are simple to
evolve and bypass the uncertainties associated with convection during
the helium burning phase. If they display the same sort of compactness
systematics seen for the full stars, one can have greater confidence
in the result.

\begin{figure}[htb]
\centering
\includegraphics[width=0.48\textwidth]{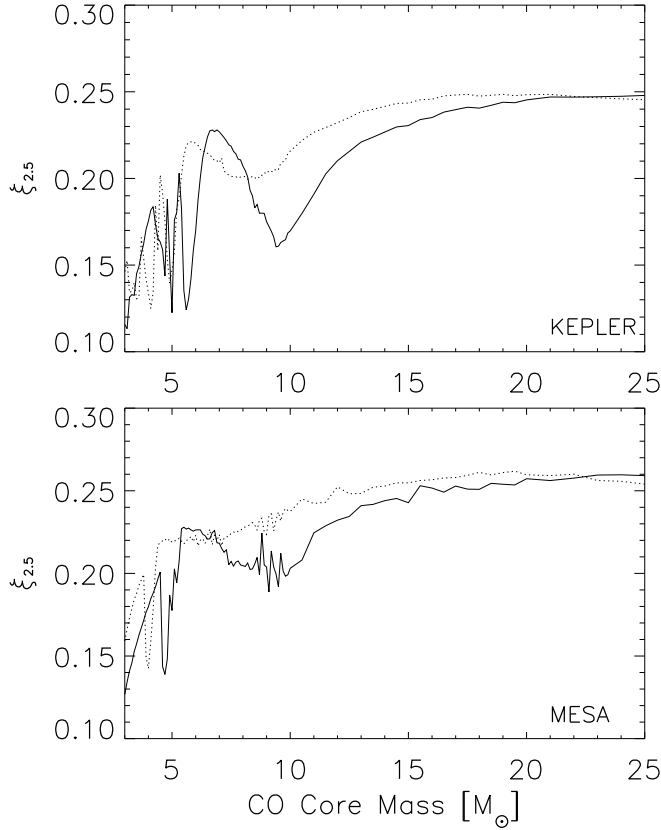}
\caption{\cp\ as a function of mass at central oxygen depletion for
  bare CO stars studied with KEPLER and MESA. Top: KS (solid) and KU
  (dotted); bottom: MS(solid) and MU(dotted). \lFig{f26}}
\end{figure}

A total of 380 CO cores were evolved using both KEPLER and MESA (Table
1).  MESA used $\alpha_{MLT}=2, \alpha_{SC}=0.1$ and $f = 0.025$ and
KEPLER used $q_r= 0.1$ and $q_{\rm Ov} = 0.01$.  The calculations
followed the evolution from central carbon ignition until the
presupernova stage for the KEPLER cores and until the core oxygen
depletion stage for MESA models. The core masses ranged from 3 to
25\Msun \ with varying increments: 0.1 \Msun \ increments were used
for 3 to 10 \Msun, 0.5 \Msun \ was used for 10 to 20 \Msun, and 1
\Msun was used for 20 to 25 \Msun. CO core masses of 3 to 9 \Msun
\ correspond, approximately, to the main sequence masses of 15 to
30\Msun \ studied in the S series. For low metallicity models (U
series) or other models with no mass loss (SH series), the maximum CO
core mass (at 65 \Msun) was 25 \Msun, and that sets the upper
limit. The physics used to calculate the CO core models was identical
to that in the full star calculations.

The different mass fractions of carbon for the S and U stars (Figure
5), however, necessitated separate surveys. Because of convection, the
composition of a CO core of given metallicity and mass can be taken as
roughly constant. A good fit to the abundances in \Fig{f5} is
\begin{equation}
\begin{split}
&X(^{12}C) = -0.037\times ln(M_{C/O})+0.221 \quad (U)\\
&X(^{12}C) = -0.035\times ln(M_{C/O})+0.253 \quad (S)
\end{split}
\end{equation}
The mass fraction of oxygen was $X(^{16}O) = 1 - X(^{12}C)$ in each
case.  These 4 sets of calculations are denoted as: KS, KU, MS and
MU., i.e. - KS stands for KEPLER cores with initial composition from
solar metallicity full stars (S Series), and MU stands for MESA cores with
initial composition from low metallicity full stars (U series).

\Fig{f26} shows \cp \ at core oxygen depletion point for the 4 series
(see also Table 1). The same sort of non-monotonic variation in \cp
\ seen previously in the full star models is evident in all cases but,
as expected, cores with initial composition from U series (KU and MU),
have their first peak at a lower mass than the cores with initial
composition from solar metallicity stars (KS and MS). This same sort
of offset was also seen in the \cp\ curve for full stars of different
metallicities (\Fig{f3}).

\begin{figure}[htb]
\centering
\includegraphics[width=0.48\textwidth]{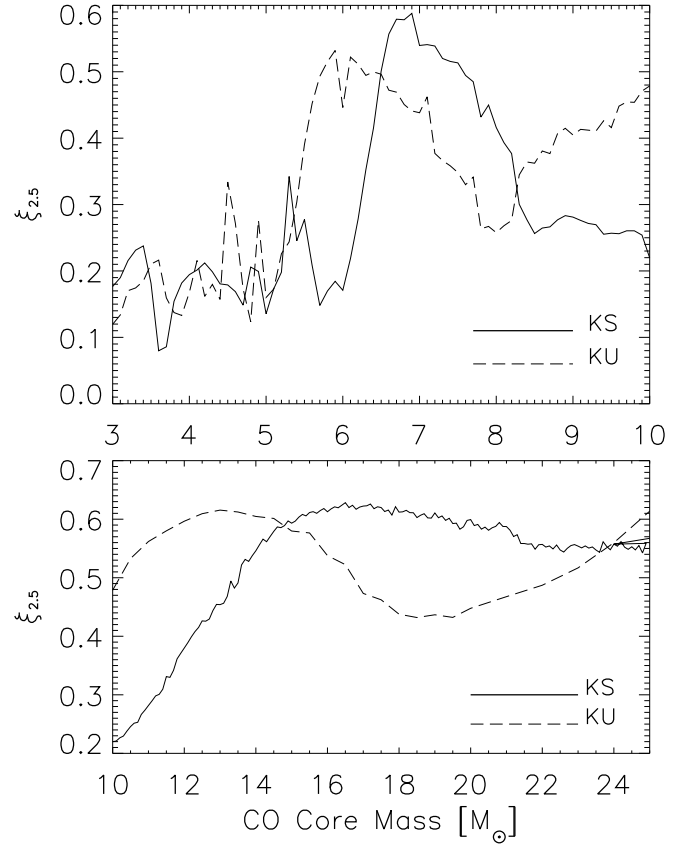}
\caption{Final \cp \ for presupernova stars derived from bare CO cores
  of a given mass and composition. (Top:) For the mass range 3 to 12
  \Msun, corresponding to main sequence masses from about 15 \Msun
  \ to 30 \Msun. Note the large bump near 6 \Msun \ corresponding
  to the similar peak for main sequence masses 22 to 24 \Msun shown in
  \Fig{f3}. The offset between solar (KS; green) and low
  metallicity (KU; red) is also similar. (Bottom:) A similar plot for
  the heavier CO cores. The agreement with \Fig{f3} is
  striking.\lFig{f27}}
\end{figure}

Above CO core masses of 11 \Msun, the differences between runs using
KEPLER and MESA are minimal. Below that, one sees similar structures
for KS and MS, for example, but there are differences. The major peak
for the KS stars at 7 \Msun, is near 5.5\Msun \ for the MS stars and
the peak of KU stars is also shifted down about 1 \Msun \ in the MU
stars. Even after removing the major dependence on CO core mass by
running bare CO cores, there is still some residual difference
resulting from the sensitivity of the location of carbon burning
shells to convective and semiconvective physics in the two codes. The
overall similarity of the patterns is encouraging, however. The
compactness is non-monotonic with mass and there are regions of mass
that will easier to explode than others.

The KEPLER calculations of CO cores were continued until the
presupernova stage, and \Fig{f27} shows the compactness parameter as a
function of core mass for both the KS and KU series at this final
stage. A comparison with \Fig{f3} is interesting. The peaks at 22 and
23 \Msun \ in the full star models for the U and S series are reflected
well in their equivalent mass CO cores, 5.8 \Msun \ and 6.2 \Msun,
respectively.  The good agreement continues to heavier masses where
the dip at 50 \Msun \ for the U-series appears as a dip at 18.5 \Msun
\ in the KU-series. The mass of the CO core in the U-model 50 \Msun
\ star was indeed 18.5 \Msun. Similarly, although not so precisely,
the CO core of the SH-series 55 \Msun \ star was 20.6 \Msun, close to
the dip at 24 \Msun \ seen in the lower panel of \Fig{f27}. The
maximum mass CO core in the SH series was 27 \Msun \ for the 65 \Msun
\ model. For the U series, the corresponding CO core mass was 26
\Msun.

\section{Conclusions}
\lSect{s6}

While characterizing the complex structure of a presupernova massive
star with a single number is a gross over-simplification, considerable
insight can be achieved by studying the systematics of the compactness
parameter, \cp.  The compactness of a given presupernova core turns
out to be sensitive to a variety of inputs, including not only the
star's initial mass and composition, but the way convection is handled
in the code, the nuclear physics employed, and the code used for the
calculation. Stellar modeling, at least of massive stars, is, in a
sense, a statistical science. Different groups will almost universally
obtain different results for the structure and composition of a
non-rotating presupernova star resulting from a given main sequence
mass star, even without the complication of mass loss. However, for
the same input physics, there should exist a robust pattern of results
that can be sampled by running a large set of models. This is a paper
about one such pattern, the compactness parameter.

Our study confirms the non-monotonic nature of \cp\ noted in previous
works, and explores its causes.  \Fig{f7} shows that the choice of 2.5
\Msun \ as a fiducial point for evaluating the compactness, though
arbitrary, is a good indicator of the behavior at other locations deep
inside the star, but outside the iron core. Whether \cp \ is evaluated
at the ``presupernova model ($V_{\rm collapse} = 1000$ km s$^{-1}$),
or at core bounce matters little at 2.5 \Msun, but for points deeper
in, small variations seen in \cp \ for the presupernova model are
amplified during the collapse. Small variations in main sequence mass
might therefore result in significant differences in outcome for an
explosion model \citep{Ugl12}.

Extensive new surveys of presupernova evolution are reported for both
solar metallicity and low metallicity (10$^{-4}$ Z$_{\rm sun}$) stars
(\Sect{s2}; Table 1). While the emphasis here is on presupernova
structure and compactness, this is clearly a rich data set for
exploring nucleosynthesis and supernova explosion physics, and this
will be done elsewhere. Major differences are found in the
presupernova compactness for stars of the same initial mass, but
different metallicity (\Sect{s2.2}). The most obvious differences result
from the differing degrees of mass loss expected for the two
compositions, but there are subtle secondary effects involving the
amount of helium convected into the helium core near central helium
depletion.  The growth of the helium convection zone results from a
different structure at its outer edge which, in turn, is affected by
the strength of the hydrogen shell, whether the star is a red or blue
supergiant, and how much hydrogenic envelope remains following mass
loss (in the solar metallicity stars). Even with mass loss suppressed,
a substantial offset remains between presupernova stars of solar and
low metallicity. Generally speaking, the low metallicity stars become
difficult to blow up at a lower mass (\Fig{f3}) and may make more
black holes.

The underlying cause for the non-monotonic behavior of \cp \ with
initial mass derives from the interaction of carbon-burning and
oxygen-burning shells with the carbon-depleted and, later, oxygen
depleted core (\Sect{s3}). As has been noted previously, the end of
convective carbon core burning at around 20 \Msun \ for standard
stellar physics, has a major effect, but not so much due to the
entropy decrease during central carbon burning itself as to the effect
this has on the ensuing shells. Above 20 \Msun, both central carbon
burning and what was the first convective shell switch to radiative
transport.  Energy generation in excess of pair neutrino losses is
unable to support core against further contraction, and the next two
carbon convective shells are pulled down. Their gradual migration
outwards again causes the rise and decline, and rise again of \cp
\ above 20 \Msun.  Above 30 \Msun, \cp\ rises to large values and, in
the absence of mass loss, keeps on rising. There is an appreciable dip
at about 50 \Msun, though, due to the outward migration of the oxygen
shell which essentially plays the same role as the carbon shells did
for lower mass stars. 

Broadly speaking, then, the evolution of massive stars separates into
four mass intervals characterized by similar values of \cp. For a
standard choice of physics and solar metallicity these are stars from
8 to 22 \Msun, 22 - 25 \Msun, 25 - 30 \Msun, and above 30 \Msun (e.g.,
\Fig{f3}). Stars below 22 \Msun \ and between 25 and 30 \Msun \ have
low \cp \ and might more easily explode. Between 22 and 25 \Msun and
above 30 \Msun, it will be more difficult for the traditional
neutrino-transport model to succeed.  There can be large variations
within those ranges and not all stars lighter than 22 \Msun \ will be
easy to explode \citep{Ugl12}, but the distribution of stars that make
black holes, for example, or important contributions to
nucleosynthesis could be bi-modal. Galactic chemical evolution studies
using such non-monotonic yields have yet to be done, but may give
quite different results, e.g., for the s-process and oxygen synthesis
\citep{Bro13}.

While the pattern seen in \cp \ {\sl vs} mass is robust, the locations
of its peaks and valleys are sensitive to the treatment of
semiconvection and convective overshoot in the code. These cause
variations in the mass of the CO core and the carbon mass fraction
when carbon burning ignites. Reducing semiconvection or convective
overshoot makes the CO core smaller and this is a leading cause of the
large variations seen in the literature for many studies of
presupernova (\Tab{cocore}). Smaller CO cores can mimic the effect of
reducing the main sequence mass in a calculation where the convection
physics is held constant and thus make cores with steep surrounding density 
gradients up to a higher mass. However these cores, while easier to blow
up will be essentially nucleosynthetically barren. Extremely small
values for semiconvective efficiency and overshoot mixing can lead to
a discrepancy between the lightest stars expected to explode as
supernovae based upon observations, about 8 \Msun, and that calculated
by theory. Caution must be exercised when using the CO core mass as a
discriminant, however, because the outcome is also very sensitive to
the carbon mass fraction, and even to the treatment of convection after
helium depletion.

Most of the models in this paper were calculated using the KEPLER code,
but the MESA code was also used to calculate the compactness curve at
oxygen deletion where its essential features have already been
determined. Reasonable agreement is achieved between the two codes
provided that some amount of semiconvection and convective overshoot
mixing is used in both. The non-monotonic nature of the compactness
plot, in particular, is apparent in the results of both surveys.

The effects of varying the uncertain reaction rate for
$^{12}$C($\alpha,\gamma)^{16}$O within its error range were also
explored and found to have an appreciable effect. A total error of just
10\% in the combined error for 3 $\alpha$ and
$^{12}$C($\alpha,\gamma)^{16}$O can shift features in the compactness
plot by a solar mass or more. To the extent the compactness is related
to the difficulty of exploding a given model using neutrinos, the
changes for nucleosynthesis and supernova ``cut-off mass'' are
significant at that level.

\begin{figure}[htb]
\centering
\includegraphics[width=0.48\textwidth]{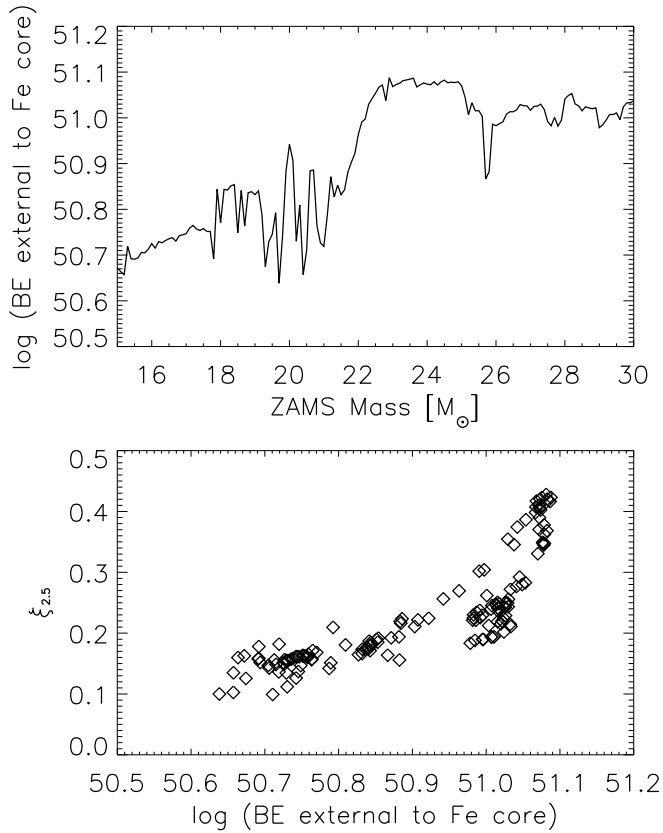}
\caption{Final \cp\ and binding energies for presupernova stars of the
  S series (Top:) The binding energy above the iron core shown as a
  function of initial mass. Despite the general increasing baseline
  trend, the relative variations closely follow that of \cp. (Bottom:)
  The final value of \cp\ is plotted against the binding energy outside of
  the iron core. \lFig{f28}}
\end{figure}

Because the CO core mass plays such a major role in determining \cp,
we explored its systematics using a grid of ``bare'' CO-cores. These
differ from the CO cores of massive stars which are defined by
composition changes and do not have sharp declines in density and
pressure at their edges.  Nevertheless, the same non-monotonic
compactness curve results. This both confirms the robustness of the
\cp \ relation found in the other surveys and suggests that, for a
given presupernova model, its final CO core mass, not its main
sequence mass might be a more accurate structural indicator.  The same
behavior was found for CO cores studied using MESA. Picking the CO
core mass circumvents some, though certainly not all of
the uncertainty introduced by ambiguous treatments of semiconvection
and convective overshoot mixing during core helium burning. For WR
stars and other stars that have experienced severe mass loss, the
presupernova CO core mass will be the best indicator of presupernova
structure.  Even for bare CO cores, however, the mass fraction of
carbon at carbon ignition remains an important ``hidden variable''
that will be sensitive to both the code physics and the metallicity of
the star.

While this paper has emphasized the relation between compactness of
the core and its explodability in terms of the neutrino transport
model, the results are actually more general. \Fig{f28} shows the
binding energy of the matter outside of the iron core of the
presupernova star as a function of its compactness. Stars with larger
values of \cp \ will be more difficult to explode by {\sl any}
mechanism, both because of the higher ram pressure immediately after
collapse and the larger energy needed to eject the matter outside the
iron core.

\section{Acknowledgements}

We gratefully acknowledge numerous valuable discussions with Alexander
Heger and Bill Paxton regarding the KEPLER and MESA codes and their
application to this problem. We also benefited from Alex's expert
insights into massive stellar evolution on many occasions, and his
permission to include unpublished details of the ``U series'' low
metallicity models.  Casey Meakin and Evan O'Connor provided
helpful comments on an earlier draft. We thank the
  anonymous referee for a careful reading of the submitted manuscript
  and many helpful suggestions. This research has been supported by
  the National Science Foundation (AST 0909129), the NASA Theory
  Program (NNX09AK36G), and the University of California Lab Fees
  Research Program (12-LR-237070).

\newpage

\begin{deluxetable}{cccccc}
\tablewidth{0pt} \tablecaption{Calculations} \tablehead{
  \colhead{Code} & \colhead{series} & \colhead{evolution} &
  \colhead{mass-range} & \colhead{\# of models} & \colhead{comments}}
\startdata

KEPLER & S & ZAMS-$5\times10^{11}$g cm$^{-3}$ & 15-30 & 151 & default\\
\nodata & SH & ZAMS-preSN & 30-60 & 9 & no mass-loss\\
\nodata & U & ZAMS-preSN & 15-65 & 86 & $10^{-4}Z_{\odot}$, no mass-loss\\
\nodata & SS & ZAMS-preSN & 15-30 & 176 & weak semiconvection\\
\nodata & SO & ZAMS-preSN & 15-30 & 16 & no overshooting\\
\nodata & SB & ZAMS-preSN & 19-25 & 65 & varied $^{12}C(\alpha,\gamma)^{16}O$\\
\nodata & KS & C.ign-preSN & 3-25 & 95 & CO core with S composition\\
\nodata & KU & C.ign-preSN & 3-25 & 95 & CO core with U composition\\

\hline\\

MESA & M & ZAMS-O.dep & 15-30 & 16 & default\\
\nodata & Mov & ZAMS-O.dep & 15-30 & 16 & with overshooting\\
\nodata & MS & C.ign-O.dep & 3-25 & 95 & CO core with S composition\\
\nodata & MU & C.ign-O.dep & 3-25 & 95 & CO core with U composition\\

\enddata

\end{deluxetable}

\begin{deluxetable}{ccccccccccc}
\tablewidth{0pc} \tablecaption{He Core and CO Core Sizes} \tablehead{
\multicolumn{2}{c}{15\Msun} & \multicolumn{2}{c}{20\Msun} &\multicolumn{2}{c}{25\Msun} & \multicolumn{2}{c}{boundary criteria} &\colhead{measured} &\colhead{comments} & \colhead{References}\\
\colhead{$M_{\alpha}$} & \colhead{$M_{CO}$} & \colhead{$M_{\alpha}$} & \colhead{$M_{CO}$} & \colhead{$M_{\alpha}$} & \colhead{$M_{CO}$} & \colhead{$M_{\alpha}$} & \colhead{$M_{CO}$} &\colhead{at\footnotemark[1]}& & }
\startdata
4.84 &2.97 & 7.17 &4.94 & 9.65 &7.20 & $X_H=0.01$ &$X_{He}=0.01$ (?) & C.dep &$\alpha_{ov}=0.2$ & \citet{Maeder92}\\
4.0\footnotemark[2] & 2.02 & 6.0\footnotemark[2] & 3.70 & 8.0\footnotemark[2] & 5.75 & \multicolumn{2}{c}{$max(\epsilon_{nuc})$ in shells} & O.dep &$\alpha_{ov}=0.0$ &\citet{Thiel96} \\
4.21 &2.44 & 6.27 &4.13 & 8.50 &6.27 & $X_{He}=0.75$ &$X_{He}=0.01$ & Si.dep &$\alpha_{ov}=0.1$ & \citet{Hirschi04}\\
5.68 &3.76 & 8.65 &6.59 & 10.0 &8.63 & $X_{He}=0.75$ &$X_{He}=0.01$ & Si.dep &and rotation & \citet{Hirschi04}\\
3.85 &2.19 & 5.70\footnotemark[3] &3.54 & 5.18 & 7.73 & $X_{H}=0.001$ &$X_{He}=0.001$ & O.dep & & \citet{Eid04}\\
4.97 &2.56 & 7.31 &3.83 & 9.80 &5.48 & \multicolumn{2}{c}{$max(\epsilon_{nuc})$ in shells} & preSN &$\alpha_{ov}=0.2,\alpha_{sc}=0.02$ & \citet{Chieffi13}\\
\vspace{0.5cm}
5.37 &3.59 & WR\footnotemark[4] &5.53 & WR\footnotemark[4] &6.63 & \multicolumn{2}{c}{$max(\epsilon_{nuc})$ in shells} &preSN & and rotation & \citet{Chieffi13}\\
4.35 &3.14 & 6.21 &4.93 & 8.25 &6.95 & $X_H=0.2$ &$X_{He}=0.2$ &preSN & $q_r=0.1, q_{ov}=0.01$   & KEPLER,S \\
4.44 &3.14 & 7.18 &5.61 & 9.07 &7.35 & $X_H=0.2$ &$X_{He}=0.2$ &preSN & and rotation            & KEPLER \\
4.29 &3.08 & 6.27 &4.99 & 8.50 &7.18 & $X_H=0.2$ &$X_{He}=0.2$ &preSN & $q_r=0.05,q_{ov}=0.01$   & KEPLER, SS\\
4.15 &2.93 & 6.22 &4.59 & 8.40 &6.39 & $X_H=0.2$ &$X_{He}=0.2$ &preSN & $q_r=0.015,q_{ov}=0.01$  & KEPLER, SS\\
4.09 &1.96 & 6.00 &2.95 & 8.30 &4.40 & $X_H=0.2$ &$X_{He}=0.2$ &preSN & $q_r=0.005,q_{ov}=0.01$  & KEPLER, SS\\
3.98 &1.63 & 6.08 &2.71 & 8.42 &4.11 & $X_H=0.2$ &$X_{He}=0.2$ &preSN & $q_r=0.001,q_{ov}=0.01$  & KEPLER, SS\\
   -   & -     & 6.22 &4.96 & 8.28 &7.00 & $X_H=0.2$ &$X_{He}=0.2$ &preSN & 1.5$\times$Buchmann      & KEPLER, SB\\
   -   & -     & 6.21 &4.90 & 8.26 &6.98 & $X_H=0.2$ &$X_{He}=0.2$ &preSN & 1.3$\times$Buchmann      & KEPLER, SB\\
   -   & -     & 6.17 &4.96 & 8.23 &6.99 & $X_H=0.2$ &$X_{He}=0.2$ &preSN & 0.9$\times$Buchmann      & KEPLER, SB\\
   -   & -     & 6.16 &4.94 & 8.21 &6.99 & $X_H=0.2$ &$X_{He}=0.2$ &preSN & 0.7$\times$Buchmann      & KEPLER, SB\\
\vspace{0.5cm}
4.19 &2.70 & 6.11 &4.38 & 8.14 &6.27 & $X_H=0.2$ &$X_{He}=0.2$ & preSN & $q_r=0.1,q_{ov}=0.0$ & KEPLER, SO\\
3.77 &1.96 & 5.81 &3.47 & 7.31 &3.91 & $X_H=0.2$ &$X_{He}=0.2$ &O.dep & $f=0.0$  & MESA, M\\
4.01 &2.21 & -      & -     & 7.80 &5.16 & $X_H=0.2$ &$X_{He}=0.2$ &O.dep & $f=0.005$  & MESA\\
4.20 &2.39 & -      & -     & 8.06 &5.47 & $X_H=0.2$ &$X_{He}=0.2$ &O.dep & $f=0.01$  & MESA\\
4.41 &2.61 & -      & -     & 8.34 &5.81 & $X_H=0.2$ &$X_{He}=0.2$ &O.dep & $f=0.015$  & MESA\\
4.85 &3.05 & 6.92 &4.77 & 9.28 &6.84 & $X_H=0.2$ &$X_{He}=0.2$ &O.dep & $f=0.025$ & MESA, Mov\\
\enddata
\tablecomments{$q_r$-Eq.3; $q_{Ov}$-Eq.5; $\alpha_{sc}$-Eq.7;
  $f$-Eq.8; $\alpha_{ov}$ - is the multiplier on the pressure scale
  height.}
\footnotetext[1]{Approximate points in the evolution where the
  core sizes were measured. The CO core size
  increases very slightly after C.dep, and almost not at all after O.dep.}
\footnotetext[2]{Calculations for bare helium cores.}
\footnotetext[3]{Averaged from 5 different values with varying
  reaction rates and convection physics for the 25\Msun \ model.}
\footnotetext[4]{Lost all of its hydrogenic envelope.}
\lTab{cocore}
\end{deluxetable}

\clearpage
\newpage

\begin{deluxetable}{cccc}
\tablewidth{0pt} 
\tablecaption{X($^{12}$C) and $^{12}C(\alpha,\gamma)^{16}O$} 
\tablehead{
  \colhead{Multiplier\footnotemark[1]} & \colhead{19 \Msun} & \colhead{22 \Msun} &
  \colhead{25 \Msun} }
\startdata

1.9 & 0.132 & 0.120 & 0.110\\
1.5 & 0.183 & 0.171 & 0.161\\
1.3 & 0.212 & 0.200 & 0.193\\
1.2 & 0.229 & 0.218 & 0.209\\
0.9 & 0.286 & 0.276 & 0.270\\
0.7 & 0.335 & 0.325 & 0.320\\

\enddata \tablecomments{Evaluated at central helium
  depletion for the KEPLER SB series.}

\footnotetext[1]{Multipliers times the \citet{Buc96,Buc97} rate.}

\end{deluxetable}

\begin{deluxetable}{ccc}
\tablewidth{0pt} 
\tablecaption{Critical Masses for Central Carbon Burning} 
\tablehead{
  \colhead{Series} & \colhead{ZAMS Mass [\Msun]} & \colhead{CO Core Mass [\Msun]} }
\startdata

S (KEPLER)   &  20.3 & 5.03 \\
U (KEPLER)   & 18.6 & 4.27 \\
Mov (MESA) & 23.0 & 5.47 \\
\vspace{0.5cm}
M (MESA)     & 28.5 & 4.86 \\

KS (KEPLER) & &5.3 \\
KU (KEPLER) & &4.6 \\
MS (MESA) & &4.6 \\
MU (MESA) & &3.9 
\enddata

\tablecomments{The CO core boundary is defined as location
  where $^4$He drops below 0.2 in mass fraction.} \lTab{critcore}
\end{deluxetable}

\clearpage

\end{document}